\pdfoutput=1

\documentclass[journal]{IEEEtran}
   \usepackage[pdftex]{graphicx}
   \graphicspath{{../pdf/}{../jpeg/}}
   \DeclareGraphicsExtensions{.pdf,.jpeg,.png}
\usepackage[caption=false,font=footnotesize]{subfig}
\usepackage{multirow}
\usepackage{makecell}
\begin{document}
\title{Textually Guided Ranking Network for Attentional Image Retweet Modeling}

\author{Zhou Zhao, Hanbing Zhan, Lingtao Meng, Jun Xiao, Jun Yu, Min Yang, Fei Wu, Deng Cai%
	\thanks{Z. Zhao, H. Zhan, L. Meng, J. Xiao, F. Wu are with the College of Computer Science, Zhejiang University, Hangzhou 310027, China (e-mail: csezhaozhou@163.com; 3150104111@zju.edu.cn; lqq119119@163.com; junx@cs.zju.edu.cn; wufei@cs.zju.edu.cn); M. Yang is with the Shenzhen Institutes of Advanced Technology, Chinese Academy of Sciences (e-mail: myang@cs.hku.hk); J. Yu is with the College of Computer Science, Hangzhou Dianzi University (e-mail: yujun@hdu.edu.cn) ; D. Cai is with the State Key Lab of CAD\&CG, Zhejiang University (e-mail: dengcai@gmail.com)}%
}


\markboth{IEEE TRANSACTIONS ON MULTIMEDIA}%
{IEEE TRANSACTIONS ON MULTIMEDIA}



\maketitle

\begin{abstract}
Retweet prediction is a challenging problem in social media sites (SMS). In this paper, we study the problem of image retweet prediction in social media, which predicts the image sharing behavior that the user reposts the image tweets from their followees. Unlike previous studies, we learn user preference ranking model from their past retweeted image tweets in SMS. We first propose heterogeneous image retweet modeling network (IRM) that exploits users' past retweeted image tweets with associated contexts, their following relations in SMS and preference of their followees. We then develop a novel attentional multi-faceted ranking network learning framework with textually guided multi-modal neural networks for the proposed heterogenous IRM network to learn the joint image tweet representations and user preference representations for prediction task. The extensive experiments on a large-scale dataset from Twitter site shows that our method achieves better performance than other state-of-the-art solutions to the problem.
\end{abstract}


% \fi
\IEEEpeerreviewmaketitle

\section{Introduction}
\IEEEPARstart{M}{icroblog} services like Twitter have become important social platforms for users to share their media contents. Retweet function is usually considered to be key mechanism that enables users to repost someone else's tweets~\cite{zhang2015retweet}. In social media sites, users who follows other users are termed as "followers" and users who are followed are termed as "followees". Central problem of retweet prediction is to model tweet sharing behavior that users repost tweets along followee-follower links so that more users are informed in SMS, which has attracted considerable attention recently in~\cite{chen2016context,firdaus2016retweet,zhang2015retweet,zhang2016retweet,feng2013retweet}.
\begin{figure}[t]
	\setlength{\abovecaptionskip}{-0.2cm}
	\setlength{\belowcaptionskip}{-0.5cm}
	\centering
	\includegraphics[width=0.38\textwidth]{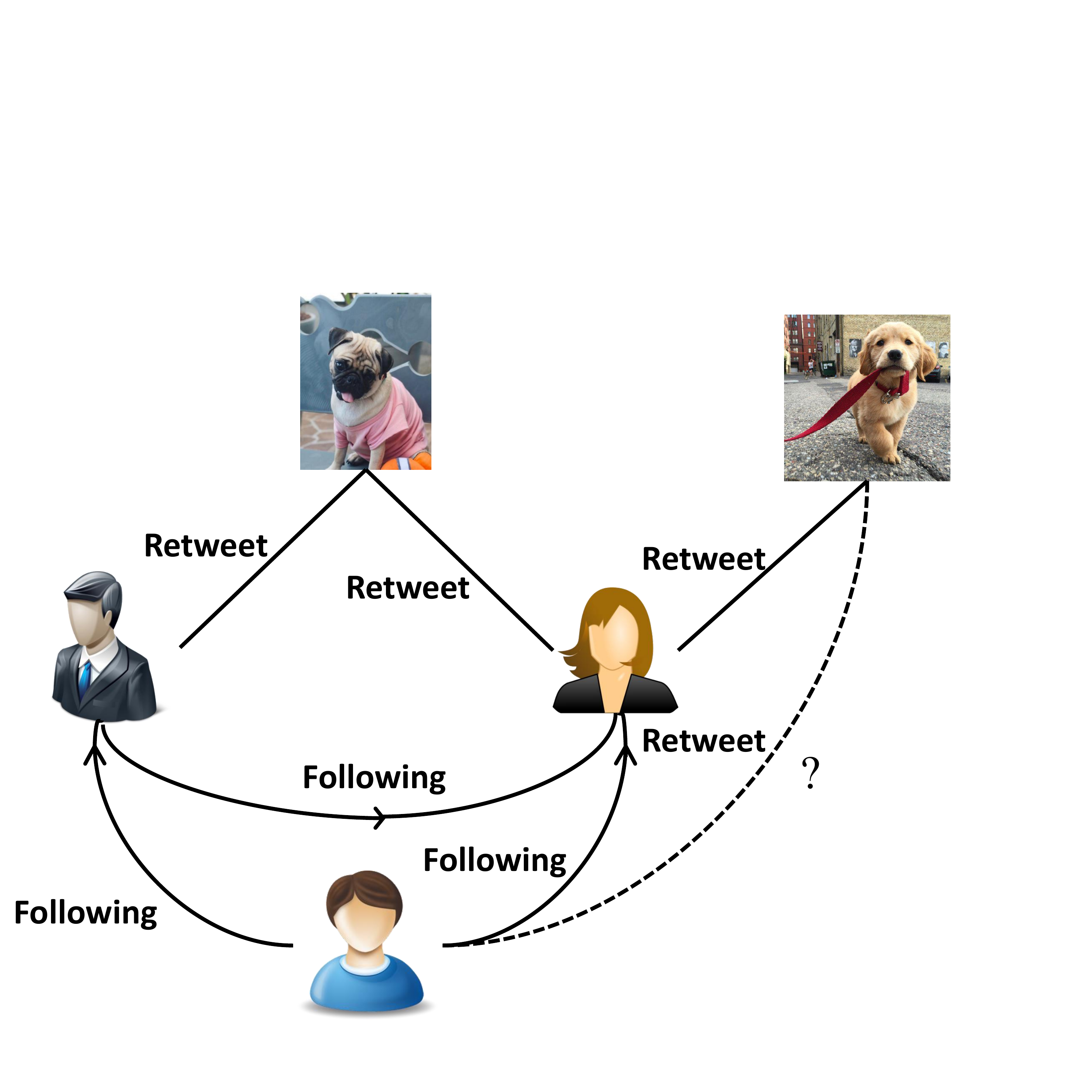}
	\caption{Example of Image Retweet Behavior. }\label{fig:demo}
\end{figure}

Existing approaches for retweet prediction~\cite{firdaus2016retweet,zhang2015retweet,zhang2016retweet,feng2013retweet,zhang2015influenced} learn user preference model from their past retweeted textual tweets, and predict users' tweet sharing behavior in SMS. With the popularity of mobile devices, the amount of user-generated image tweets grows tremendously. For example, there are about 17.2\% of tweets associated with images in Twitter~\cite{chen2016context}. So it is important to study problem of image retweet prediction in SMS. We give a simple example of image retweet prediction in Figure~\ref{fig:demo}. As there is not discriminative feature representation for tweets with image ~\cite{chen2016context} and SMS data is sparse~\cite{firdaus2016retweet} , existing proposed retweet prediction methods are ineffective to image retweet prediction problem.

Currently, most of existing retweet prediction methods~\cite{firdaus2016retweet,zhang2015retweet,zhang2016retweet,feng2013retweet,zhang2015influenced} learn semantic representation of tweet based on hand-crafted feature (e.g., bag-of-words). Recently, high-level visual features for image representation with pre-trained CNNs have shown success in various visual recognition tasks~\cite{szegedy2013deep,zhao2018social}. Since image tweets are always visual data, it is natural to employ deep convolutional neural networks~\cite{simonyan2014very} to learn visual representation of image tweets. On the other hand, image tweets are often associated with textual context information such as users' comments and captions~\cite{chen2016context}. Contextual image tweet information usually convey important messages and can gain better understanding of tweets. Since textual contextual information is always sequential data with variant length, we employ deep recurrent neural networks~\cite{hochreiter1997long} to learn its semantic representation. We employ multi-modal neural network learning method~\cite{atrey2010multimodal} to learn joint image tweet representation from their multi-modal contents, which provides complementary information with different modalities.

Sparsity of SMS data is also a challenging issue for image retweet prediction. In SMS sites, network between image tweets and users is constructed through users' retweet relations on image tweets. Usually, each user only retweets a few image tweets and thus SMS network is sparse. Inspired by homophily hypothesis~\cite{yuan2014unified}, it is possible and reasonable to assume that collective information from users' followees and users' retweeted tweets can be jointly considered for tackling the sparsity problem of image retweet prediction. It is observed that social impact for retweet behavior varies between user and his/her different followees. We thus employ attention mechanism~\cite{luong2015effective} to adaptively incorporate users' followee preference for jointly predicting targeted user's image retweet behavior.

In this paper, we study image retweet prediction problem from viewpoint of attentional multi-faceted ranking network learning. We first propose heterogeneous image retweet modeling (IRM) network that exploits multi-modal image tweets, users' retweet behaviors and their following relations for image retweet prediction. We introduce textually guided multi-modal neural networks with two sub-networks, where recurrent neural networks learn semantic representations of image tweets' contextual information, and convolutional neural networks learn visual representations. Multi-modal fusion layer is added to learn joint image tweet representation from texually guided multi-modal neural networks. We develop attentional multi-faceted ranking method with introduced multi-modal neural networks, such that multi-faceted ranking metric is implicitly embedded in user preference representation for image retweet prediction. Main contributions of this paper are summarized as follows:
\begin{itemize}
	\item Unlike previous studies, we present image retweet prediction problem from viewpoint of attentional multi-faceted ranking network learning. We propose heterogeneous IRM network to model the problem, which exploits multi-modal image tweets, users' retweet behaviors and their following relations.
	\item We develop attentional multi-faceted ranking method with textually guided multi-modal neural networks to learn user preference representation based on retweeted tweets and following relations for image tweet prediction.
	\item We evaluate our method's performance using dataset collected from Twitter. Extensive experiments show that our method outperforms several state-of-the-art solutions to the problem.
\end{itemize}

The rest of this paper is organized as follows. In Section~\ref{sec:model}, we present the problem of image retweet prediction from the viewpoint of attentional multi-faceted ranking network learning. Many experimental results are presented in Section~\ref{sec:experiments}. We provide a brief review of the related work about retweet prediction in Section~\ref{sec:related work}. Finally, we provide some concluding remarks in Section~\ref{sec:conclusion}.

\begin{figure*}[t]
	\setlength{\abovecaptionskip}{-0.8cm}
	\setlength{\belowcaptionskip}{-0.0cm}
	\centering
	\includegraphics[width=1\textwidth]{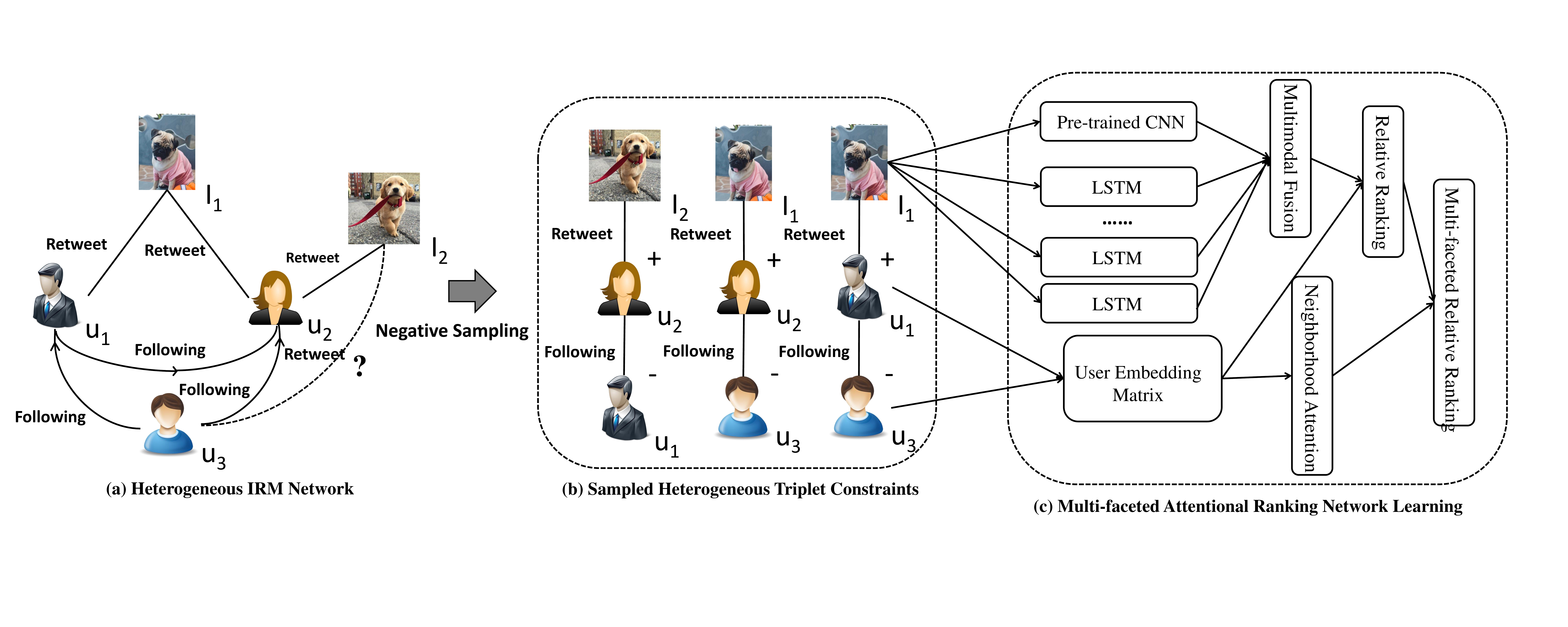}
	\caption{The Overview of Attentional Multi-faceted Ranking Network Learning for Image Retweet Prediction. (a) The heterogeneous IRM network is constructed by integrating multi-modal image tweets, users' past retweet behaviors and their following relations. (b) A negative sampling based method is employed on the heterogeneous IRM network to sample the relative users' preference. (c) The attentional multi-faceted ranking network learning method is invoked with multi-modal neural networks based on relative user preference loss for image retweet prediction. }\label{fig:framework}
\end{figure*}

\section{Image Retweet Prediction via Attentional Ranking Network Learning}\label{sec:model}

In this section, we first present the problem of image retweet prediction from the viewpoint of heterogeneous image retweet modeling network learning. We then propose the attentional multi-faceted ranking method based on social impact of the relative followee preference. We devise the textually guided multi-modal network to guide the image region through the user's contextual attention, thus jointly representing the image tweet and its captions or comments. 

\subsection{The Problem}
Before presenting the problem, we first introduce some basic notions and terminologies. Since image tweets are always visual data, it is natural to employ deep convolutional neural networks~\cite{simonyan2014very} to learn visual representation of image tweets. Given a set of image tweets $I=\{{\bf i}_{1},{\bf i}_{2},\ldots,{\bf i}_{n}\}$, we first learn the image tweets' convolutional feature by the pretrained CNN's last convolutional layer as $X=\{{\bf x}_{1},{\bf x}_{2},\ldots,{\bf x}_{n}\}$, where ${\bf x}_{i}$ is a 3-dimension feature containing both the location and the visual information of image. We also learn the image's visual embedding by the same convolutional neural networks' last fully connected layer as $F=\{{\bf f}_{1},{\bf f}_{2},\ldots,{\bf f}_{n}\}$. In the next section we introduce how to use the visual embedding feature to guide the location of image's convolutional feature.  On the other hand, textual context information of image tweets such as users' comments and captions also gain better understanding of image tweets. We thus employ deep recurrent neural networks~\cite{hochreiter1997long} to learn its semantic representation. Given a set of textual contexts $D=\{{\bf d}_{1},{\bf d}_{2},\ldots,{\bf d}_{n}\}$, we take recurrent neural networks' last hidden layer as semantic embedding of textual contexts by $Y=\{{\bf y}_{1},{\bf y}_{2},\ldots,{\bf y}_{n}\}$, where ${\bf y}_{i}=\{{\bf y}_{i1},{\bf y}_{i2},\ldots,{\bf y}_{ik}\}$ denotes semantic embeddings of the image tweet's different captions and comments. We denote the joint image tweet representations by $Z=\{{\bf z}_{1},{\bf z}_{2},\ldots,{\bf z}_{n}\}$, where ${\bf z}_{i}$ is joint representation of the $i$-th image tweet based on its visual representation ${\bf x}_{i}$ and contextual semantic representation ${\bf y}_{i}$. We denote the set of ranking models for user preference representation by $U=\{{\bf u}_{1},{\bf u}_{2},\ldots,{\bf u}_{m}\}$, where ${\bf u}_{j}$ is preference representation embedding of the $j$-th user.

Recently, existing approaches for retweet prediction~\cite{firdaus2016retweet,zhang2015retweet,zhang2016retweet,feng2013retweet,zhang2015influenced} learn user preference model from their past retweeted textual tweets, and then predict users' tweet sharing behavior. Unlike previous studies, we propose attentional multi-faceted ranking metric heterogeneous IRM (i.e., image retweet modeling) network that exploits multi-modal image tweets, users' past retweet behaviors and their following relations for image retweet prediction. We denote proposed heterogeneous IRM network by $ G=(V, E)$, where the set of nodes $V$ is composed of the joint image tweet representations $ Z $ and user preference representations $ U $, the set of edges $ E $ consists of users' past retweeted behaviors $ H $ and their following relations $ S $. We denote the retweeted behaviors between image tweets and users by matrix $ H \in R^{n\times m}$, where the entry $h_{ij}=1$ if the $i$-th image tweet is retweeted by the $j$-th user, otherwise, $h_{ij}=0$. We then consider the following relations between users by matrix $ S \in R^{m\times m}$, where $s_{ij}=1$ if the $i$-th user follows the $j$-th user. We next denote the set of the $i$-th user's followees by $N_{i}$ (i.e., ${\bf u}_{j}\in N_{i}$ if $s_{ij}=1$), and the total set of users' followees by $N=\{N_{1},N_{2},\ldots,N_{m}\}$. We illustrate a simple example of the heterogeneous IRM network in Figure~\ref{fig:framework}(a).

We then derive the heterogeneous triplet constraints from the IRM network as the users' relative preference for training the attentional multi-faceted ranking networks. We consider that the users express the explicit positive interest on the image tweets when he/she retweeted them in the IRM networks. On the other hand, following the existing Twitter analysis works~\cite{chen2012collaborative}, we consider that the users may show the implicit negative interest on the non-retweeted image tweets of their followees. This is because the non-retweeted image tweets by the followees are more likely to be seen but disliked by the user.

Given retweeted behavior between the $i$-th image tweet ${\bf z}_{i}$ and the $j$-th user ${\bf u}_{j}$ (i.e., $h_{ij}=1$), we sample a non-retweeted image tweet of ${\bf u}_{j}$'s followees as ${\bf z}_{k}$. Following popular homophily hypothesis~\cite{yuan2014unified}, we also incorporate users' followee preference for image tweet modeling. We then model users' relative preference by ordered tuple $(j,i,k,N_{j})$, meaning that \lq\lq the $j$-th user prefers the $i$-th image tweet to the $k$-th one\rq\rq. Let $T=\{(j,i,k,N_{j})\}$ denote set of ordered tuples obtained from IRM network for a set of $n$ image tweets and $m$ users. We then consider ordered heterogeneous tuples as the constraints for learning user preference representations. More formally, we aim to learn the multi-faceted ranking metric function for image retweet prediction. For any $(j,i,k,N_{j})\in T$, the inequality holds:
\begin{small}
	\begin{eqnarray}
	F_{{\bf u}_{j}}({\bf z}_{i}) > F_{{\bf u}_{j}}({\bf z}_{k}) \Longleftrightarrow f_{{\bf u}_{j}}({\bf z}_{i})h_{{\bf N}_{j}}({\bf z}_{i}) > f_{{\bf u}_{j}}({\bf z}_{k})h_{{\bf N}_{j}}({\bf z}_{k}) ,\nonumber
	\end{eqnarray}
\end{small}
where $F_{{\bf u}_{j}}(\cdot)=f_{{\bf u}_{j}}(\cdot)h_{{\bf N}_{j}}(\cdot)$ is the multi-faceted ranking model of the $j$-th user for image retweet prediction. The function $f_{{\bf u}_{j}}(\cdot)$ is the personalized ranking model of the $j$-th user and $h_{{\bf N}_{j}}(\cdot)$ models the social impact of the relative followee preference on the $j$-th user. We then define the personalized ranking function by $f_{{\bf u}_{j}}({\bf z}_{i})={\bf u}_{j}^{T}{\bf z}_{i}$, where ${\bf u}_{j}$ is the relative preference of the $j$-th user and ${\bf z}_{i}$ is the joint representation of the $i$-th image tweet. We will present the details of the function $h_{{\bf N}_{j}}(\cdot)$ in the next section.

Using the notations above, we define the problem of image retweet prediction from the viewpoint of attentional multi-faceted ranking network learning as follows. Given the input image tweets $I$ with their associated contexts $D$, the set of ordered tuples for users' relative preference $T$, and the heterogeneous IRM network $G$, our goal is to learn the multi-faceted ranking metric representations for all user preferences $U$ and the multimodal image tweet contents $Z$, and then rank the image tweets for the targeted users for image retweet prediction. The image tweets to user ${\bf u}$ are then ranked according to the multi-faceted user preference function $F_{{\bf u}}(\cdot)$.

\subsection{Attentional Textually Guided Ranking Network Learning}

In this section, we propose the attentional multi-faceted ranking network with the textually guided multi-modal layer for image retweet prediction. We present the learning process in Figures~\ref{fig:framework}(a),~\ref{fig:framework}(b) and~\ref{fig:framework}(c).

We first choose proper multi-modal neural networks for image tweet representation in IRM networks, which consists of two sub-networks: a deep convolutional neural network for visual representation of image data, and a deep recurrent neural network for semantic representation of textual contextual data. These two sub-networks interact with each other in a multi-modal fusion layer to form the joint representation, illustrated in Figures~\ref{fig:framework}(b) and~\ref{fig:framework}(c). For the visual representation of the image data, we use the activation of the last convolutional layer and last fully connected layer of the proposed convolutional neural network Inception Net~\cite{Szegedy2015GoingDW}, which has been widely used in many visual representation tasks~\cite{zhang2017relation,zhao2017deep,zhao2017videodual}. Meanwhile, we train the LSTM networks~\cite{hochreiter1997long} for the associated contexts of image tweet, and then take the output the last LSTM cell as its semantic representation. Considering the fact that the associated context of image tweets may be in the paragraph of several sentences with user comments and captions, we split them into sentences to learn the semantic representations by LSTM networks.

In order to learn the joint representation of image tweets with different modalities, a simple way is to set up a linear sum multi-modal layer that connects the textual representation oriented from recurrent neural network part and visual representation oriented from convolutional neural network part. For different textual representation oriented from recurrent neural network part ${\bf y}_{i}=\{{\bf y}_{i1},{\bf y}_{i2},\ldots,{\bf y}_{ik}\}$, we fuse them by an additional max-pooling layer. We then map the activation of the two layers (i.e., the visual representation of image tweets and the semantic representation of textual contexts) into the same multi-modal feature fusion space and add them together to obtain the activation of the multi-modal fusion layer, given by

\begin{figure}[t]
	\setlength{\abovecaptionskip}{-0.2cm}
	\setlength{\belowcaptionskip}{-0.5cm}
	%\centering
	\hspace{-35mm}
	%\vspace{20mm}
	\includegraphics[width=1.17\textwidth]{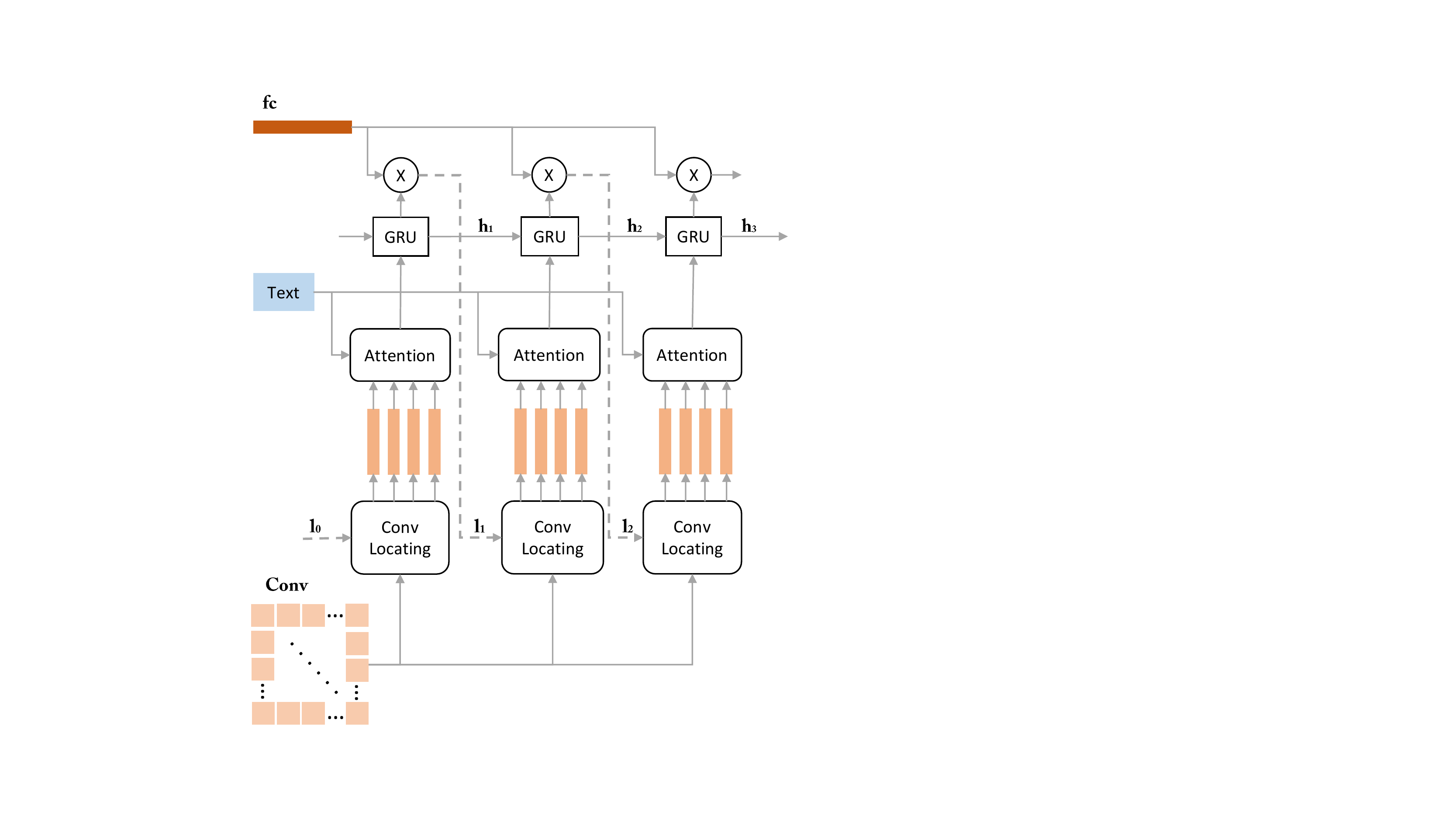}
	\vspace{-10mm}
	\caption{Textually guided multi-modal fusion network }\label{fig:multimodal}
\end{figure}

\begin{eqnarray*}
	{\bf z}_{i} = g({\bf W}^{(i)}{\bf f}_{i}+{\bf W}^{(d)}{\bf y}_{i}),
\end{eqnarray*}

where $+$ denotes the element-wise addition for the next location representation with different modalities. The matrix ${\bf W}^{(i)}$ and ${\bf W}^{(d)}$ are weight matrices. The $g(\cdot)$ is the element-wise scaled hyperbolic tangent function, which forces the gradients into the most non-linear value range and leads to a faster training process, proposed in~\cite{lecun2012efficient}.

However, such simple method doesn't take advantage of the contextual relation between different comments and their matched image tweets. In order to get a more relevant representation of image tweeets and textual comments, we set up the textually guided multi-modal fusion layer that connects the textual representation oriented from recurrent neural network part and visual representation oriented from convolutional neural network part, illustrated in Figure~\ref{fig:multimodal}. Because each image tweet have many captions and comments from its publisher and subscribers, we suppose that different comments express both associated and extended information of image. Therefore, instead of using the visual feature from the last fully connected layer of pretrained CNN, we use the image's convolutional feature which contains both the location and visual feature of image to generate the appropriate representation of users' focus on the image tweet. 

In order to locate the image's proper region for the user's focus, we denote the location mapping vector by $L=\{{\bf l}_{0},{\bf l}_{1},\ldots,{\bf l}_{k}\}$, where ${\bf l}_{i}= \{{\bf l}_{xi},{\bf l}_{yi}\}$ represents the x-axis and y-axis coordinate in the image convolutional feature respectively. Given the convolutional feature ${\bf x}_{i}$ and the location mapping vector ${\bf l}_j$, the conv locating in Figure~\ref{fig:multimodal} extracts a multi-dimensional feature $\eta({\bf x}_{i},{\bf l}_j)$ from ${\bf x}_{i}$ centered at ${\bf l}_j$. We then fuse the textual embedding with our extracted convolutional feature using the attention mechanism. Given the semantic representation of  $j$-th comment of $i$-th image ${\bf y}_{ij}$ and the multi-dimensional feature $\eta({\bf x}_{i},{\bf l}_j)=\{\eta_{i1},\eta_{i2},\ldots, \eta_{ik}\}$, the textual attention score for the $j$-th comment and the $k$-th convolutional feature is given by 

\begin{eqnarray*}
	s_{jk} = {\bf p}\cdot tanh ({\bf W}^{(t)}{\bf y}_{ij}+{\bf W}^{(u)}\eta_{ik}+{\bf b}),
\end{eqnarray*}

where ${\bf W}^{(t)}$ and ${\bf W}^{(u)}$ are parameter matrices. The ${\bf b}$ is the bias vector and ${\bf p}$ is the parameter vector for computing the textual attention score. For each followee $\eta_{k}$ in $\eta({\bf x}_{i},{\bf l}_j)$, its score activation is given by $\alpha_{k}=\frac{\exp(s_{jk})}{\sum_{\eta_k\in \eta({\bf x}_{i},{\bf l}_j)}\exp(s_{jk})}$. Thus, the textual impact on the $j$-th image convolutional feature is given by $g_{ij}=\sum_{k}\alpha_{k}\eta_{ik}$.

In order to get the high-level representation of our attentional image feature which is combined with the textual information, we use another recurrent neural network to infer the location of next image region. With $g_{ij}$ as the input of $j$-th time step, the RNN's hidden state and output are denoted by ${\bf h}_{ij}$ and ${\bf c}_{ij}$. The visual feature here from pretrained CNN's last fully connected layer is taken as the image's global information to facilitate the locating process. Given the image's visual embedding ${\bf f}_i$ and the RNN's $j$-th step's output ${\bf c}_{ij}$, the next location mapping vector is given by 

\begin{eqnarray*}
	{\bf l}_{j+1} = g({\bf W}^{(j)}{\bf f}_{i}+{\bf W}^{(c)}{\bf c}_{ij}),
\end{eqnarray*}
where $+$ denotes the element-wise addition with different modalities. The matrix ${\bf W}^{(j)}$ and ${\bf W}^{(c)}$ are weight matrices. The $g(\cdot)$ is the element-wise scaled hyperbolic tangent function.

We define the above described procedure as the textually guide process $G({\bf l}_{j},{\bf x}_i,{\bf y}_{ij},{\bf f}_{i})$. By stacking our model with the recurrent neural network, we can obtain the next location mapping vector and the RNN's hidden state by

\begin{figure*}[t]
	\centering
	
	\subfloat[Number of Retweets]{\includegraphics[width=0.9\columnwidth]{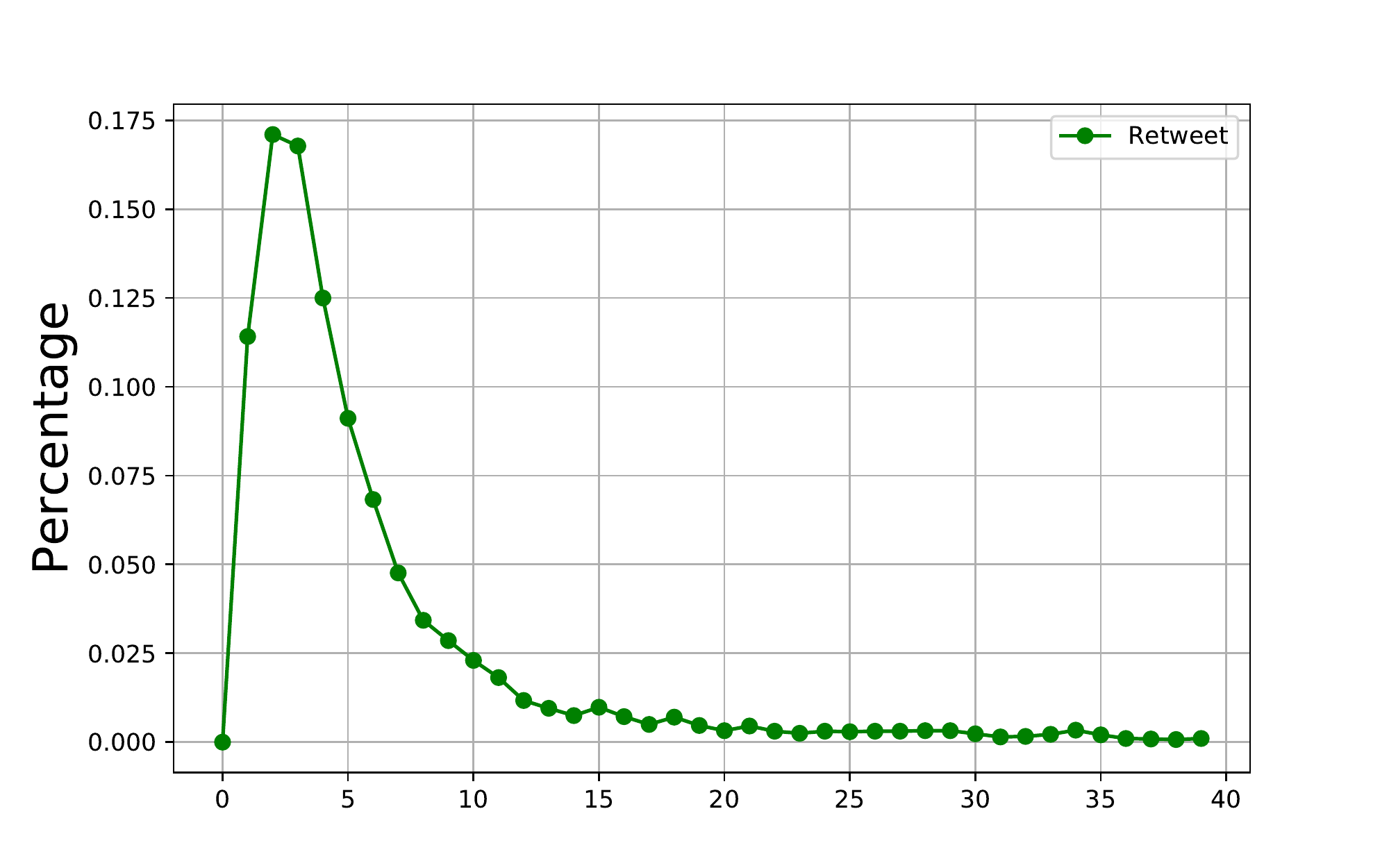}}
	\hspace{.3in}
	\subfloat[Number of Users' Follower/Followee]{\includegraphics[width=0.9\columnwidth]{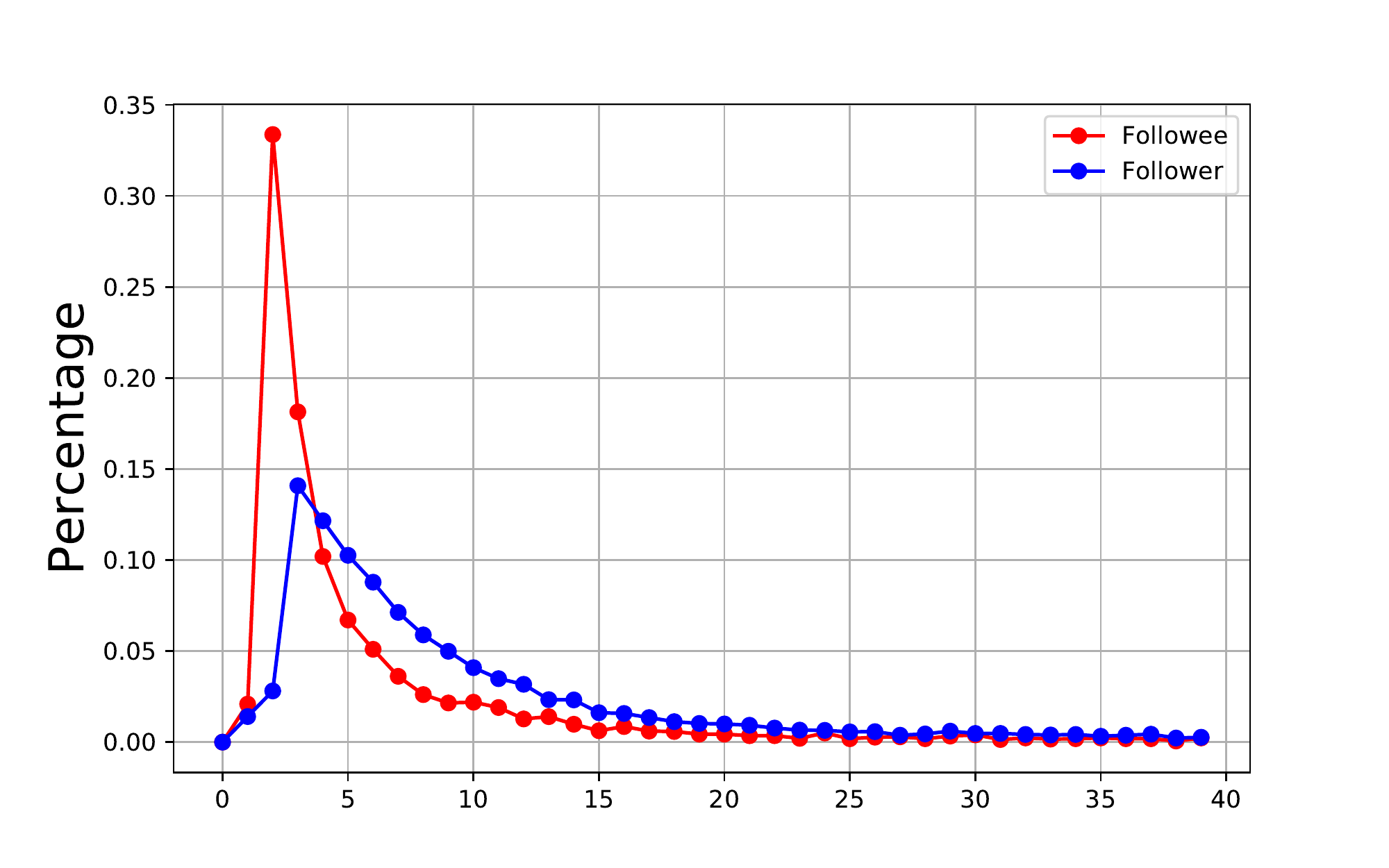}}
	\caption{The distribution of retweets and followers/followees}\label{fig:dataset}
	
\end{figure*}

\begin{eqnarray*}
	({\bf l}_{j+1},{\bf h}_{ij})=G({\bf l}_{j},{\bf x}_i,{\bf y}_{ij},{\bf f}_{i}),
\end{eqnarray*}
\begin{eqnarray*}
	{\bf z}_{ij}= W^{(c)}_j{\bf h}_{ij}
\end{eqnarray*}
where $W^{(c)}_j$ is the transformative matrice to compute the joint representation of the $i$-th image tweet. We initialize the ${\bf l}_0$ with the random strategy and obtain the last iteration's output ${\bf z}_i$ as the joint representation of the $i$-th image tweet.

We then present the attentional multi-faceted ranking function learning for image retweet prediction. Inspired by the attention mechanism~\cite{luong2015effective,zhao2017video}, we design the social impact function $h_{{\bf N}_{j}}(\cdot)$ based on the ordered tuple constraints $T=\{(j,i,k,N_{j})\}$ as follows. Given the user preference representations $U=\{{\bf u}_{1},{\bf u}_{2},\ldots,{\bf u}_{m}\}$, the social preference attention score for the $p$-th user and his/her $q$-th followee user in $N_{p}$ is given by
\begin{eqnarray*}
	s_{pq} = {\bf p}\cdot tanh ({\bf W}^{(s)}{\bf u}_{p}+{\bf W}^{(n)}{\bf u}_{q}+{\bf b}),
\end{eqnarray*}
where ${\bf W}^{(s)}$ and ${\bf W}^{(n)}$ are parameter matrices to model the preference correlation between the user and his/her followee. The ${\bf b}$ is the bias vector and ${\bf p}$ is the parameter vector for computing the social preference attention score. For each followee ${\bf u}_{q}$ in $N_{p}$, its preference activation is given by $\alpha_{q}=\frac{\exp(s_{pq})}{\sum_{q\in N_{p}}\exp(s_{pq})}$. Thus, the the social impact of the relative followee preference on the $j$-th user is given by $h_{{\bf N}_{j}}({\bf z}_{i})=\sum_{{\bf u}_{q}\in N_{j}}\alpha_{q}f_{{\bf u}_{q}}({\bf z}_{i})$.

Given the formulation of personalized ranking function $f_{{\bf u}_{j}}(\cdot)$ and social impact function $h_{N_{j}}(\cdot)$, we now design the attentional multi-faceted ranking loss function as follows:
\begin{eqnarray*}
	\mathcal{L}_{(j,i,k,N_{j})} = max(0,c+F_{{\bf u}_{j}}^{-}({\bf z}_{k})-F_{{\bf u}_{j}}^{+}({\bf z}_{i})),
\end{eqnarray*}
where the ranking function $F_{{\bf u}_{j}}({\bf z}_{i})=f_{{\bf u}_{j}}({\bf z}_{i})h_{{\bf N}_{j}}({\bf z}_{i})$, the superscript $F_{{\bf u}_{j}}^{+}(\cdot)$ indicates the positive preference and $F_{{\bf u}_{j}}^{-}(\cdot)$ denotes the negative preference. We denote the hyper-parameter $c$ ($0<c<1$) controls the margin in the loss function.

We next introduce the details of our proposed attention multi-faceted ranking network learning. We denote all the model coefficients including neural network parameter, the joint image tweet representations and user preference representation by $\Psi$. Therefore, the objective function in our learning process is given by
\begin{eqnarray*}
	\min_{\Psi}\mathcal{L}(\Psi)=\sum_{(j,i,k,N_{j})\in T}\mathcal{L}_{(j,i,k,N_{j})}(\Psi)+\beta\|\Psi\|^{2},
\end{eqnarray*}
where $\beta$ is the trade-off parameter between the training loss and regularization term. To optimize the objective, we employ the stochastic gradient descent (SGD) with diagonal variant of AdaGrad~\cite{kingma2014adam}.

\section{Experiments}\label{sec:experiments}

\subsection{Data Preparation}

\subsubsection{Information of dataset}
We collect data from Twitter, which is a popular microblog services for Web users to share their media contents~\cite{java2007we}. Users usually show their positive preference on image tweets by retweeting them in social media sites. We crawl profile of the users including their past retweeted image tweets and their following relations. In total, we collect 9,900 users, 7,193 image tweets and 29,501 following relations. We report that the average time that an image tweet retweeted by some collected users is 12.2, and the average number of image tweets that some collected user retweets is 9.1. Average number of followees among the collected users is 6.2, and maximum number of followees is 162. Average number of words in the context of image tweets is 9.1, and its standard variance is 5.4. For each retweet behavior (i.e., $h_{ij}=1$) of the user, we sample two negative image tweets from his/her followees. We sort users' retweet behaviors based on their timestamp and use the first 60\%, 70\% and 80\% of data as training set and the remaining for testing, so the training and testing data do not have overlap. The validation data is obtained separately from the training and testing data. The dataset will be released later for further study.

Figure~\ref{fig:dataset} shows the distribution of image retweets for our dataset. We can find that the number of retweet for each image is mostly within the range of 1 and 10. The distribution of all users' followees and followers are also shown in Figure~\ref{fig:dataset}, which indicates that the number of followee/follower for each user is between 3 and 7. The figure also shows a similar distribution between the number of every user's follower and followee.

\subsubsection{Image Feature Extraction}
We pre-process our collected image tweets as follows. We extract the global feature from the last fully-connected layer of pretrained Inception-V4 network for the image's feature embedding, which is the 1536-dimensional vector. To meet with the demand of our textually guided multi-modal network, we also extract the image feature from the last convolution layer of the same pretrained network, thus obtaining 8x8x1536 feature vector for each image.

\subsubsection{Text Feature Extraction}
We first filter all emoji and interjection for all captions and comments. Then for each word in sentences, we employ the pretrained Glove ~\cite{Pennington2014GloveGV} model  to extract the semantic representation. The dimension of word vector is 300. Specifically, we set four sentences for each image tweet and the length of each sentence is 12. For those image tweets which have less than 4 captions or comments, we duplicate the last comment for padding. The size of vocabulary is set to 12500 for our dataset. Therefore, we use the token  $<$unk$>$   for the out-of-vocabulary word and $<$eos$>$ to mark the end of caption or comment.

\subsection{Evaluation Criteria}

Retweet prediction task usually aims at providing top $K$ image tweets to a user in most online media services. To evaluate the effectiveness of our method in terms of top-$K$ ranked image tweets, we adopt two ranking-based evaluation criteria, Precision@K~\cite{chen2016context} and AUC~\cite{he2015vbpr,rendle2009bpr,lirelaxed} to evaluate the performance of image retweet prediction. Given test set of users $U^{t}$ and image tweets $i^{t}$, we denote predicted ranking of the top $K$ image tweets from test set for a certain user ${\bf u}_{i}$ by $R^{{\bf u}_{i}}$, where size of ranking list $|R^{{\bf u}_{i}}|$ is $K$.

\subsection{Performance Comparison}

We evaluate performance of our method {\bf AMNL} (only use linear fusion method) and {\bf AMNL+} (use the textually guided multi-modal network) with five other state-of-the-art solutions to problem of image retweet prediction as follows
\vspace{0.53em}
\begin{itemize}
	\item {\bf CITING}~\cite{chen2016context} method is the context-aware image tweet modelling framework, which explores both the image's intrinsic context and extrinsic context such as Web URL for the learning of image tweets.
	\vspace{0.5em}
	\item {\bf VBPR}~\cite{he2015vbpr} method is the scalable factorization model, which encodes the visual signal of product by deep network to predict user's feedback.
	\vspace{0.5em}
	\item {\bf FAMF}~\cite{rendle2009bpr} method is the optimization of Bayesian analysis for item recommendation, where a personalized ranking criteria and generic algorithm are designed for the item prediction task
	\vspace{0.5em}
	\item {\bf ADABPR}~\cite{rendle2014improving} method is the improvement of pairwise algorithm for recommendation systems, where a non-uniform item sampler is used to accelerate the convergence of learning network. 
	\vspace{0.5em}
	\item {\bf RRFM}~\cite{lirelaxed} method is the relaxed ranking-based factor model, which builds two-level optimization for the pairwise ranking 
	
\end{itemize}

\vspace{0.5em}

\setlength{\tabcolsep}{5pt}
\begin{table}[t]
	\makegapedcells
	\setcellgapes{3pt}
	\vspace{0.2cm}
	\footnotesize
	\centering
		\caption{Experimental results on Precision@1 with different proportions of data for training.}\label{table:precision1}
	\begin{tabular}{|c|c|c|c|}
		\hline
		\multirow{2}{*}{Method} & \multicolumn{3}{c|}{Precision@1}\\
		\cline{2-4}
		&	60\%	&	70\%	&	80\%	\\
		\hline
		RRFM	&	0.6098	&	0.6064	&	0.6261	\\
		VBPR	&	0.5914	&	0.6111	&	0.6215	\\
		FAMF	&	0.6808	&	0.6428	&	0.6071	\\
		ADABPR	&	0.6156	&	0.6134	&	0.6146	\\
		CITING	&	0.7379	&	0.71	&	0.7145	\\
		AMNL	&	0.8571	&	0.8828	&	0.8604	\\
		{\bf AMNL+} & {\bf 0.9217}	&	{\bf 0.9450}	&	{\bf 0.9411}	\\
		\hline
	\end{tabular}

	\vspace{1em}
	\centering
		\caption{Experimental results on Precision@3 with different proportions of data for training.}\label{table:precision3}
	\begin{tabular}{|c|c|c|c|}
		\hline
		\multirow{2}{*}{Method} & \multicolumn{3}{c|}{Precision@3}\\
		\cline{2-4}
		&	60\%	&	70\%	&	80\%	\\
		\hline
		RRFM	&	0.5876	&	0.6188	&	0.632	\\
		VBPR	&	0.5792	&	0.5934	&	0.6252	\\
		FAMF	&	0.5859	&	0.5297	&	0.5066	\\
		ADABPR	&	0.5703	&	0.5903	&	0.6297	\\
		CITING	&	0.7163	&	0.7044	&	0.7391	\\
		AMNL	&	0.7313	&	0.7429	&	0.7659	\\
		{\bf AMNL+} & {\bf 0.8488}	&	{\bf 0.8661}	&	{\bf 0.8627}	\\
		\hline
	\end{tabular}

	\vspace{1em}
		\caption{Experimental results on AUC with different proportions of data for training.}\label{table:auc}
	\begin{tabular}{|c|c|c|c|}
		\hline
		\multirow{2}{*}{Method} & \multicolumn{3}{c|}{AUC}\\
		\cline{2-4}
		&	60\%	&	70\%	&	80\%	\\
		\hline
		RRFM	&	0.4805	&	0.499	&	0.5051	\\
		VBPR	&	0.5118	&	0.5256	&	0.5254	\\
		FAMF	&	0.5092	&	0.5034	&	0.5078	\\
		ADABPR	&	0.5017	&	0.5008	&	0.501	\\
		CITING	&	0.5004	&	0.5067	&	0.5029	\\
		AMNL	&	0.7528	&	0.7977	&	0.8244	\\
		{\bf AMNL+} & {\bf 0.8637}	&	{\bf 0.8886}	&	{\bf 0.8968}	\\
		\hline
	\end{tabular}

	\vspace{-1.5em}
\end{table}

\setlength{\tabcolsep}{2pt}
\begin{table}[t]
	\makegapedcells
	\setcellgapes{3pt}
	\vspace{0.2cm}
	%\vspace{-0.1cm}
	\footnotesize
	\centering
		\caption{Experimental results with different modalities and components using 80\% of the data for training.}\label{table:modalities}
	
	\begin{tabular}{|c|c|c|c|}
		\hline
		Method & Precision@1 & Precision@3 & AUC\\
		\hline
		AMNL$_{i}$&	0.8039	&	0.7038	&	0.7509	\\
		AMNL$_{d}$&	0.7493	&	0.6749	&	0.7233	\\
		AMNL$_{hfunc}$& 0.8143	&	0.7299	&	0.7481	\\
		AMNL	&	0.8604	&	0.7659	&	0.8244	\\
		AMNL+$_{i}$&	0.8916	&	0.8083	&	0.8521	\\
		AMNL+$_{hfunc}$&	0.8812	&	0.8021	&	0.8373	\\
		{\bf AMNL+}	&	{\bf 0.9411}	&	{\bf 0.8627}	&	{\bf 0.8968}	\\
		\hline
	\end{tabular}

	\vspace{-1.5em}
\end{table}

\begin{figure*}[t]
	\centering
	
	\subfloat[Precision@1]{\includegraphics[width=0.7\columnwidth]{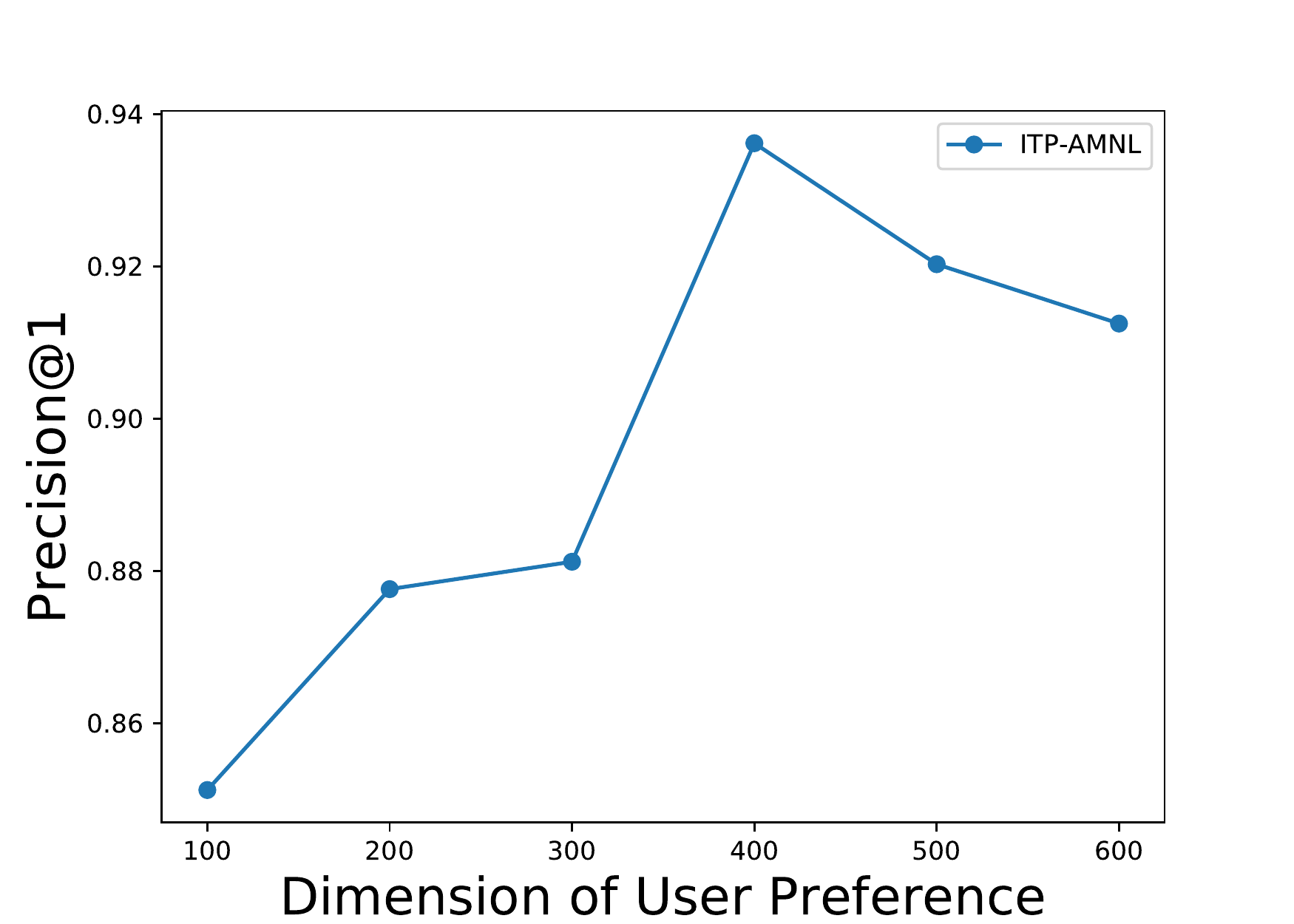}}
	\subfloat[Precision@3]{\includegraphics[width=0.7\columnwidth]{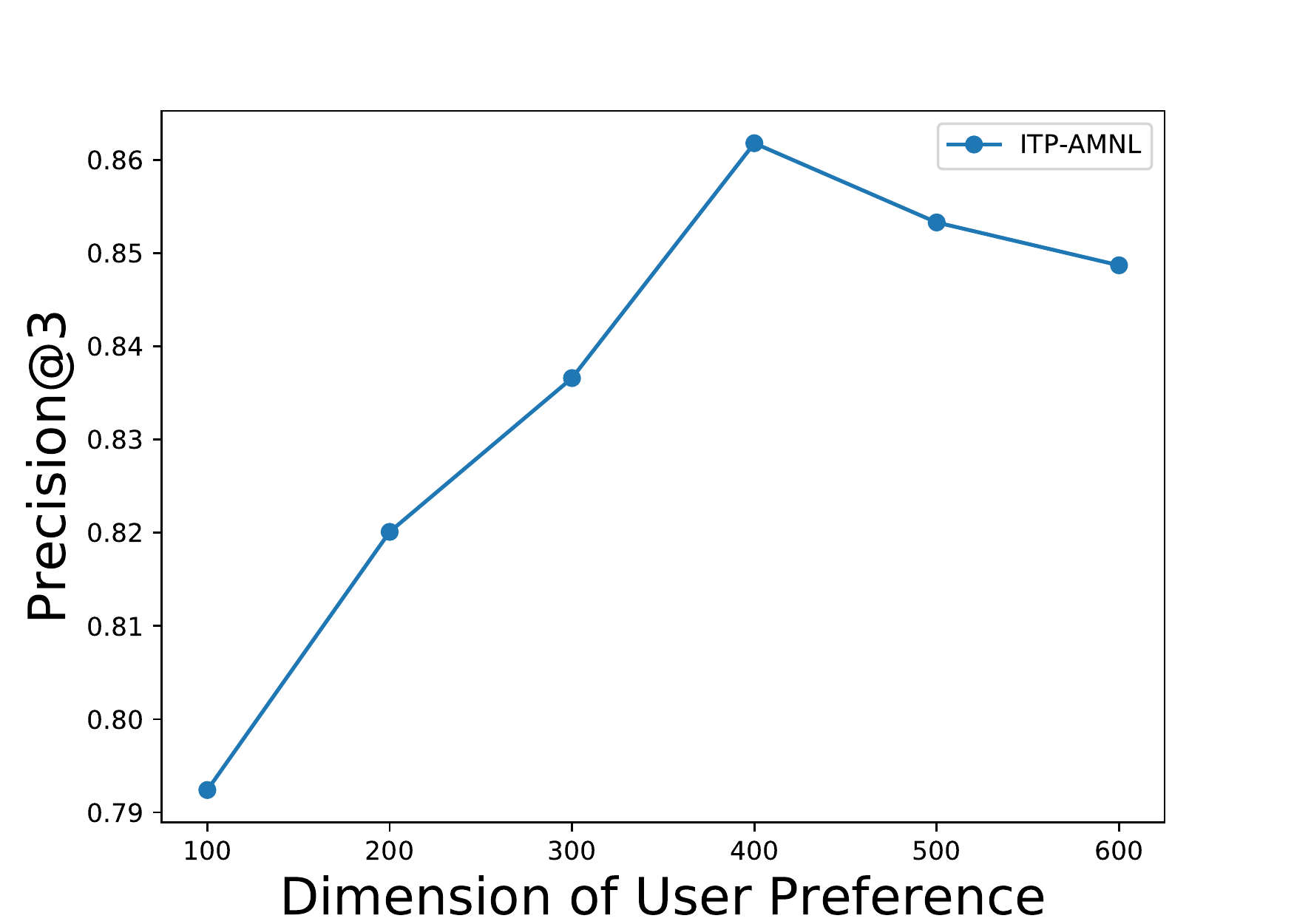}}
	\subfloat[AUC]{\includegraphics[width=0.7\columnwidth]{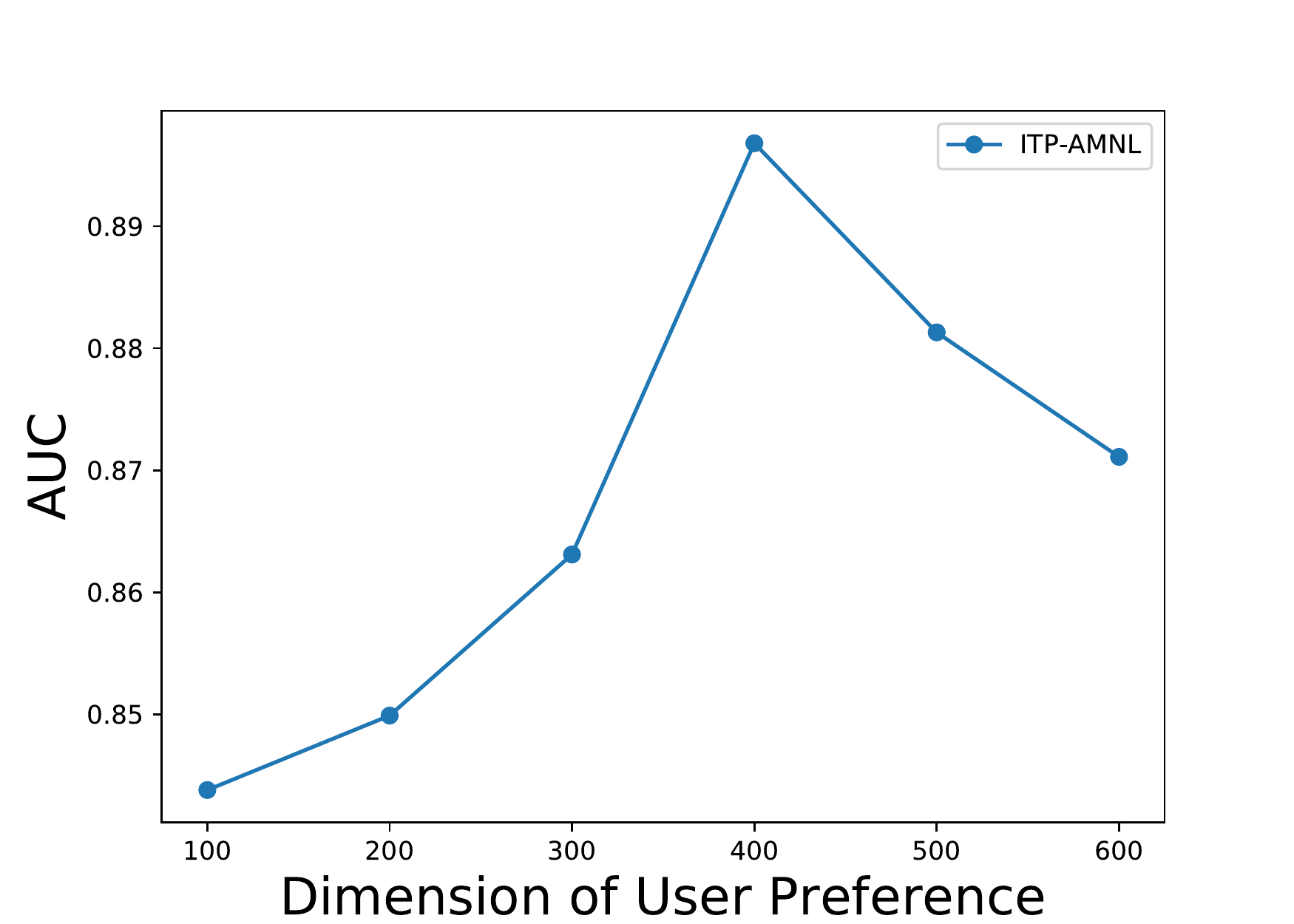}}
	\caption{Effect of the user preference dimension on Precision@1, Precision@3 and AUC using 60\% of the data for training.}\label{fig:dimension06}
	%\centering
	%\subfigure[Precision@1]{\includegraphics[width=0.6\columnwidth]{p1_07dim}}
	%\subfigure[Precision@3]{\includegraphics[width=0.6\columnwidth]{p3_07dim}}
	%\subfigure[AUC]{\includegraphics[width=0.6\columnwidth]{auc_07dim}}
	%\caption{Effect of the user preference dimension on Precision@1, Precision@3 and AUC using 70\% of the data for training.}\label{fig:dimension07}
	%\centering
	%\subfigure[Precision@1]{\includegraphics[width=0.6\columnwidth]{p1_dim}}
	%\subfigure[Precision@3]{\includegraphics[width=0.6\columnwidth]{p3_dim}}
	%\subfigure[AUC]{\includegraphics[width=0.6\columnwidth]{auc_dim}}
	%\caption{Effect of the user preference dimension on Precision@1, Precision@3 and AUC using 80\% of the data for training.}\label{fig:dimension08}
\end{figure*}

\begin{figure*}[t]
	\centering
	
	\subfloat[Precision@1]{\includegraphics[width=0.7\columnwidth]{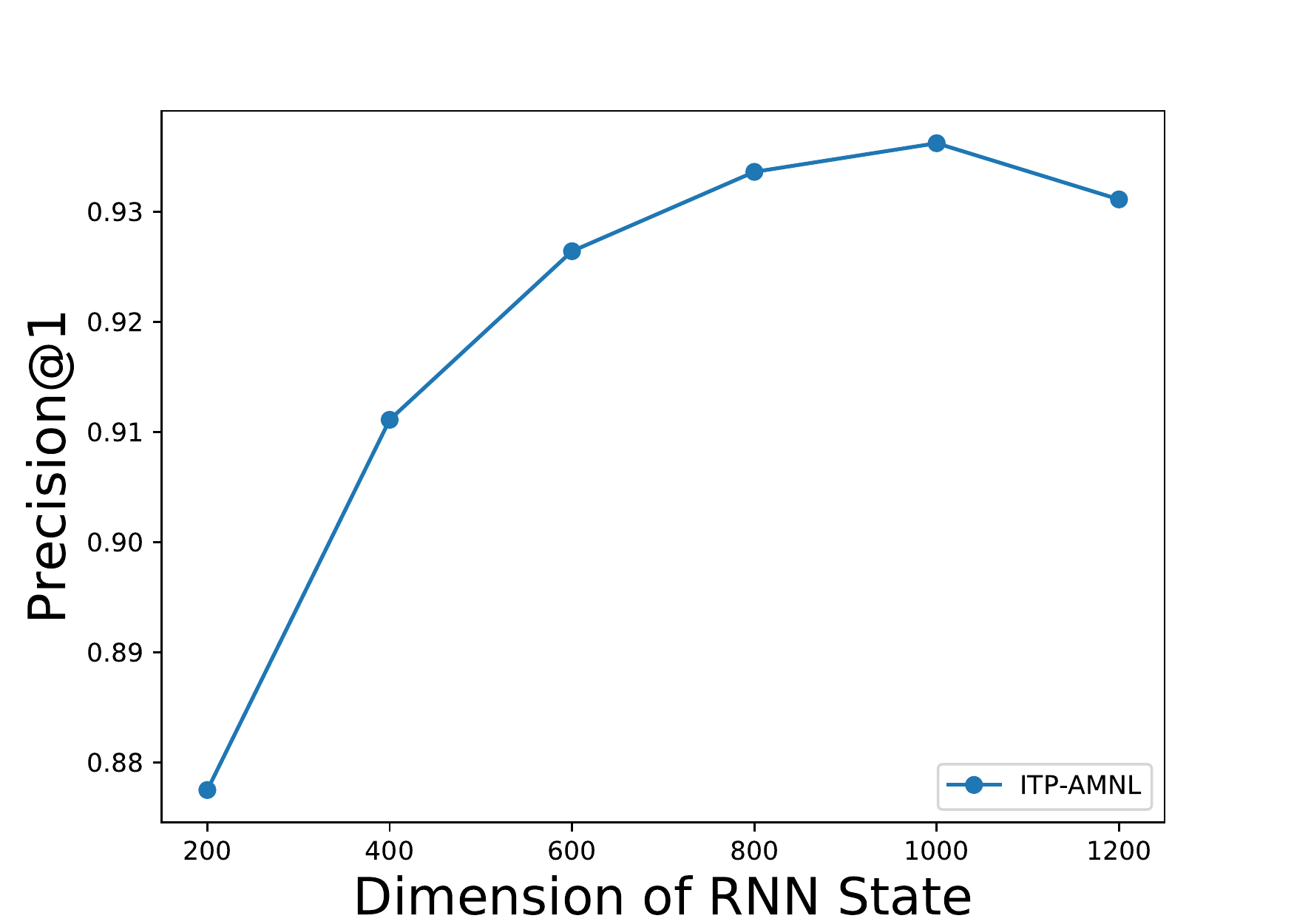}}
	\subfloat[Precision@3]{\includegraphics[width=0.7\columnwidth]{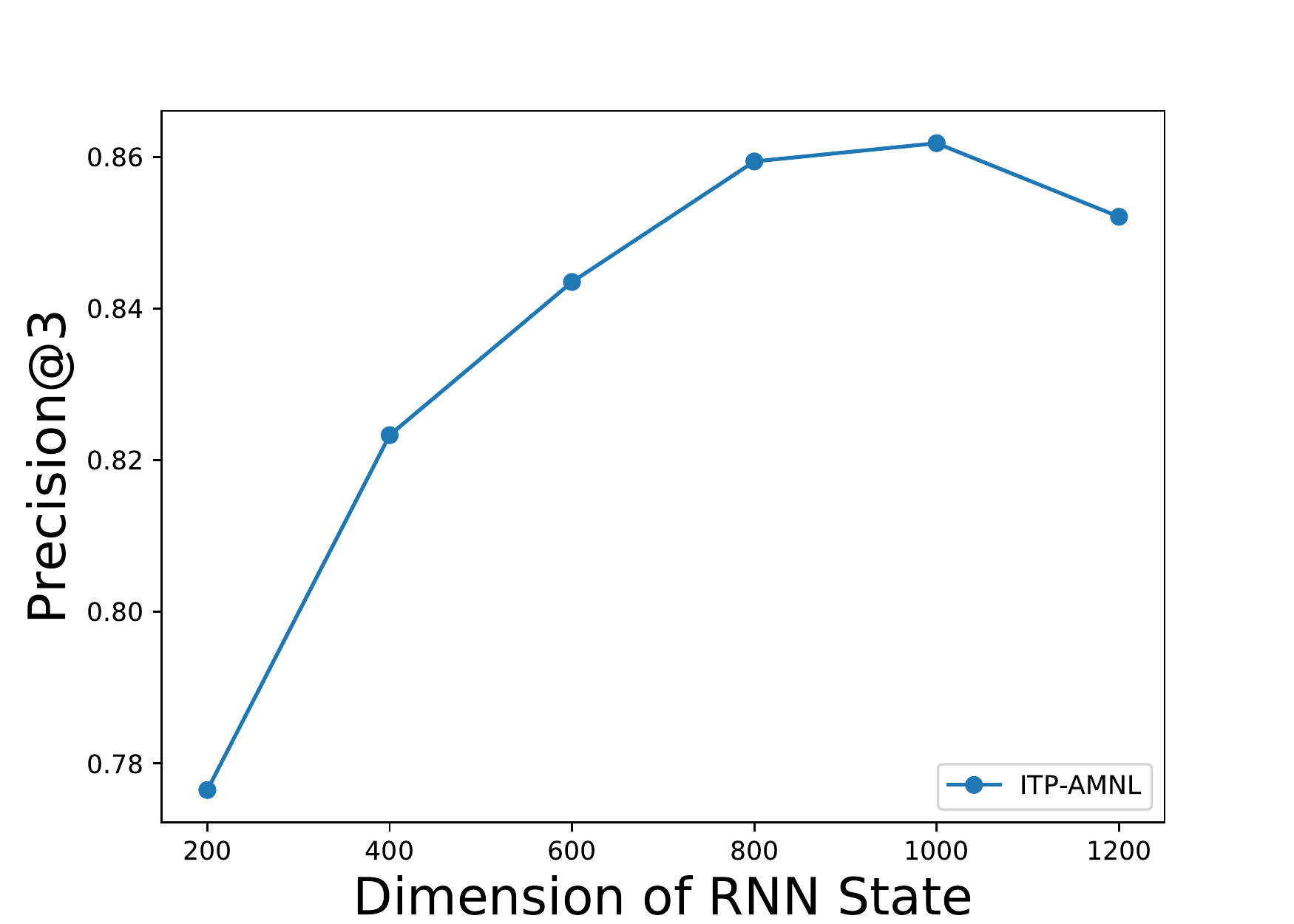}}
	\subfloat[AUC]{\includegraphics[width=0.7\columnwidth]{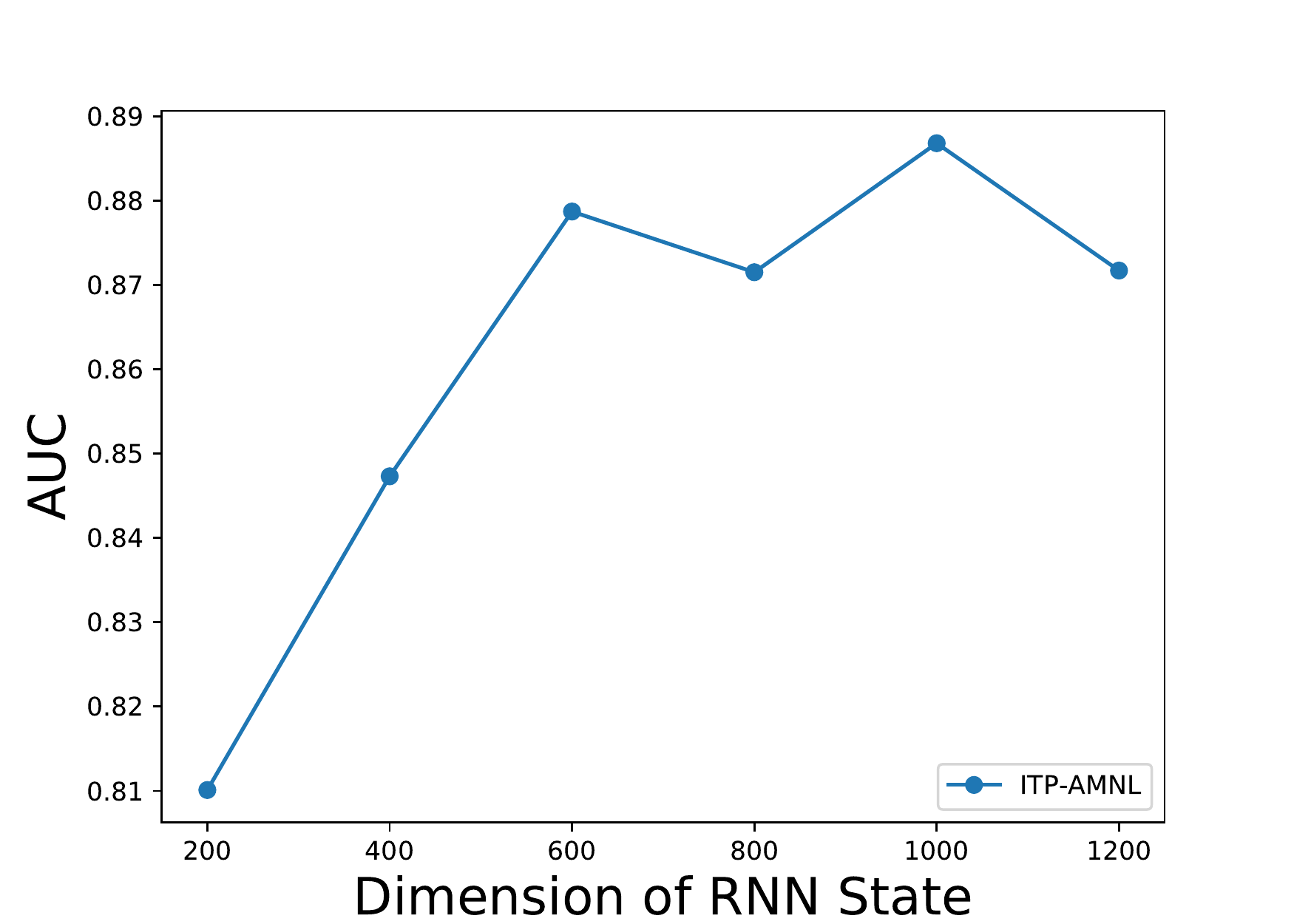}}
	\caption{Effect of the RNN state dimension on Precision@1, Precision@3 and AUC using 60\% of the data for training.}\label{fig:dimension07}
	%\centering
	%\subfigure[Precision@1]{\includegraphics[width=0.6\columnwidth]{p1_07dim}}
	%\subfigure[Precision@3]{\includegraphics[width=0.6\columnwidth]{p3_07dim}}
	%\subfigure[AUC]{\includegraphics[width=0.6\columnwidth]{auc_07dim}}
	%\caption{Effect of the user preference dimension on Precision@1, Precision@3 and AUC using 70\% of the data for training.}\label{fig:dimension07}
	%\centering
	%\subfigure[Precision@1]{\includegraphics[width=0.6\columnwidth]{p1_dim}}
	%\subfigure[Precision@3]{\includegraphics[width=0.6\columnwidth]{p3_dim}}
	%\subfigure[AUC]{\includegraphics[width=0.6\columnwidth]{auc_dim}}
	%\caption{Effect of the user preference dimension on Precision@1, Precision@3 and AUC using 80\% of the data for training.}\label{fig:dimension08}
\end{figure*}

\begin{figure*}[t]
	\centering
	
	\subfloat[Precision@1]{\includegraphics[width=0.7\columnwidth]{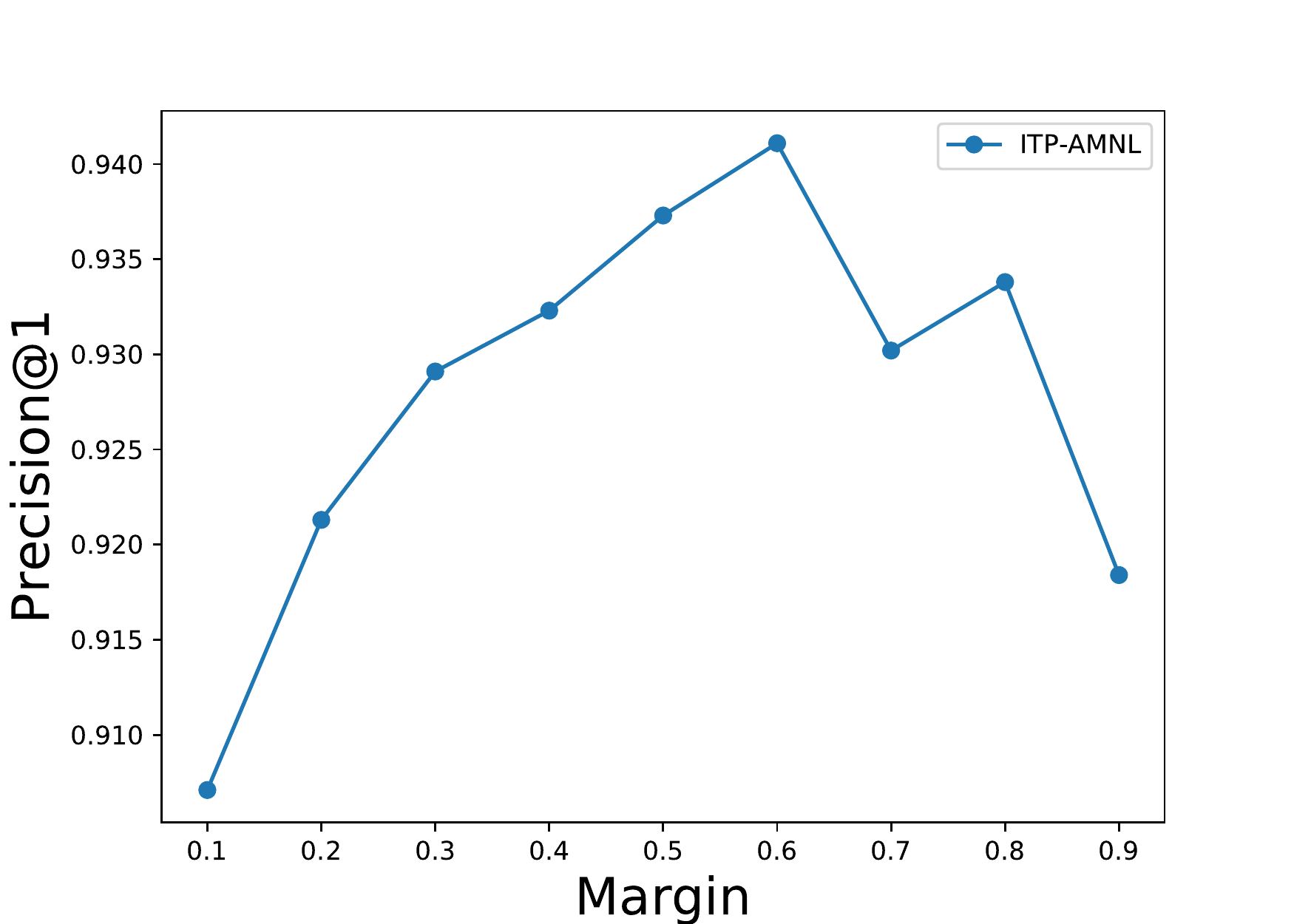}}
	\subfloat[Precision@3]{\includegraphics[width=0.7\columnwidth]{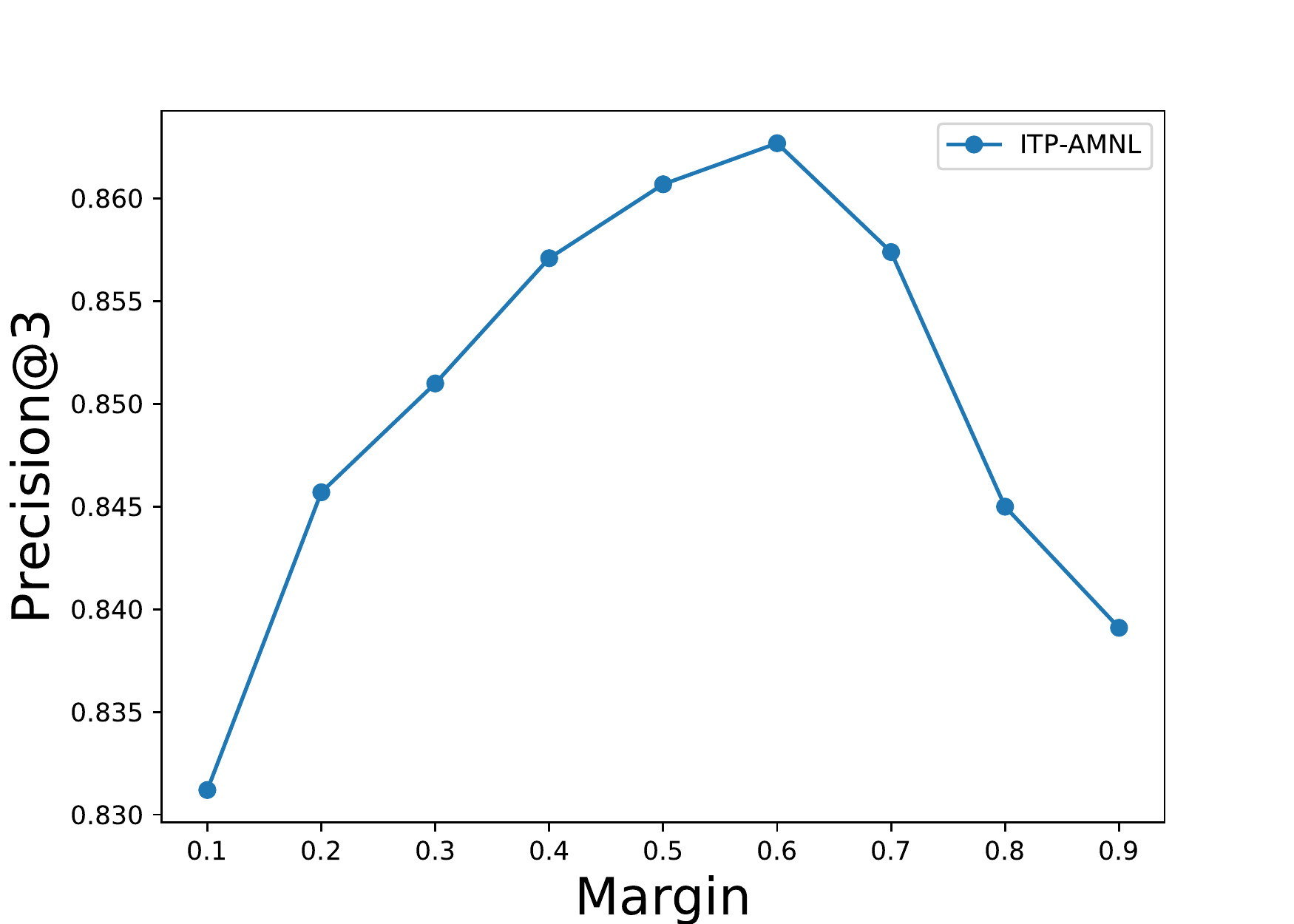}}
	\subfloat[AUC]{\includegraphics[width=0.7\columnwidth]{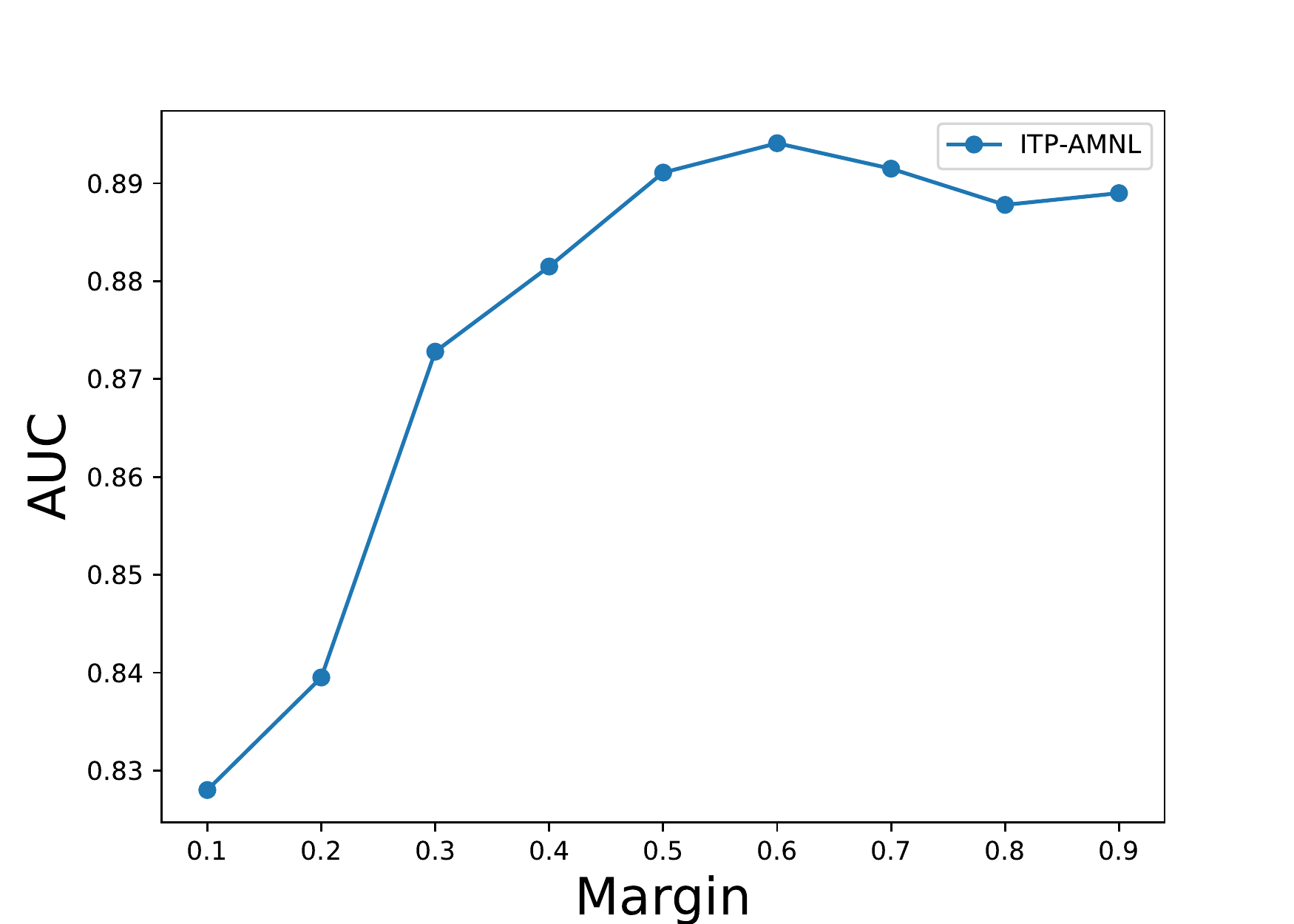}}
	\caption{Effect of the margin value on Precision@1, Precision@3 and AUC using 60\% of the data for training.}\label{fig:dimension08}
	%\centering
	%\subfigure[Precision@1]{\includegraphics[width=0.6\columnwidth]{p1_07dim}}
	%\subfigure[Precision@3]{\includegraphics[width=0.6\columnwidth]{p3_07dim}}
	%\subfigure[AUC]{\includegraphics[width=0.6\columnwidth]{auc_07dim}}
	%\caption{Effect of the user preference dimension on Precision@1, Precision@3 and AUC using 70\% of the data for training.}\label{fig:dimension07}
	%\centering
	%\subfigure[Precision@1]{\includegraphics[width=0.6\columnwidth]{p1_dim}}
	%\subfigure[Precision@3]{\includegraphics[width=0.6\columnwidth]{p3_dim}}
	%\subfigure[AUC]{\includegraphics[width=0.6\columnwidth]{auc_dim}}
	%\caption{Effect of the user preference dimension on Precision@1, Precision@3 and AUC using 80\% of the data for training.}\label{fig:dimension08}
\end{figure*}

\begin{figure*}[t]
	\setlength{\abovecaptionskip}{-0cm}
	\setlength{\belowcaptionskip}{-0.45cm}
	\centering
	\subfloat[Objective value]{\includegraphics[width=0.8\columnwidth]{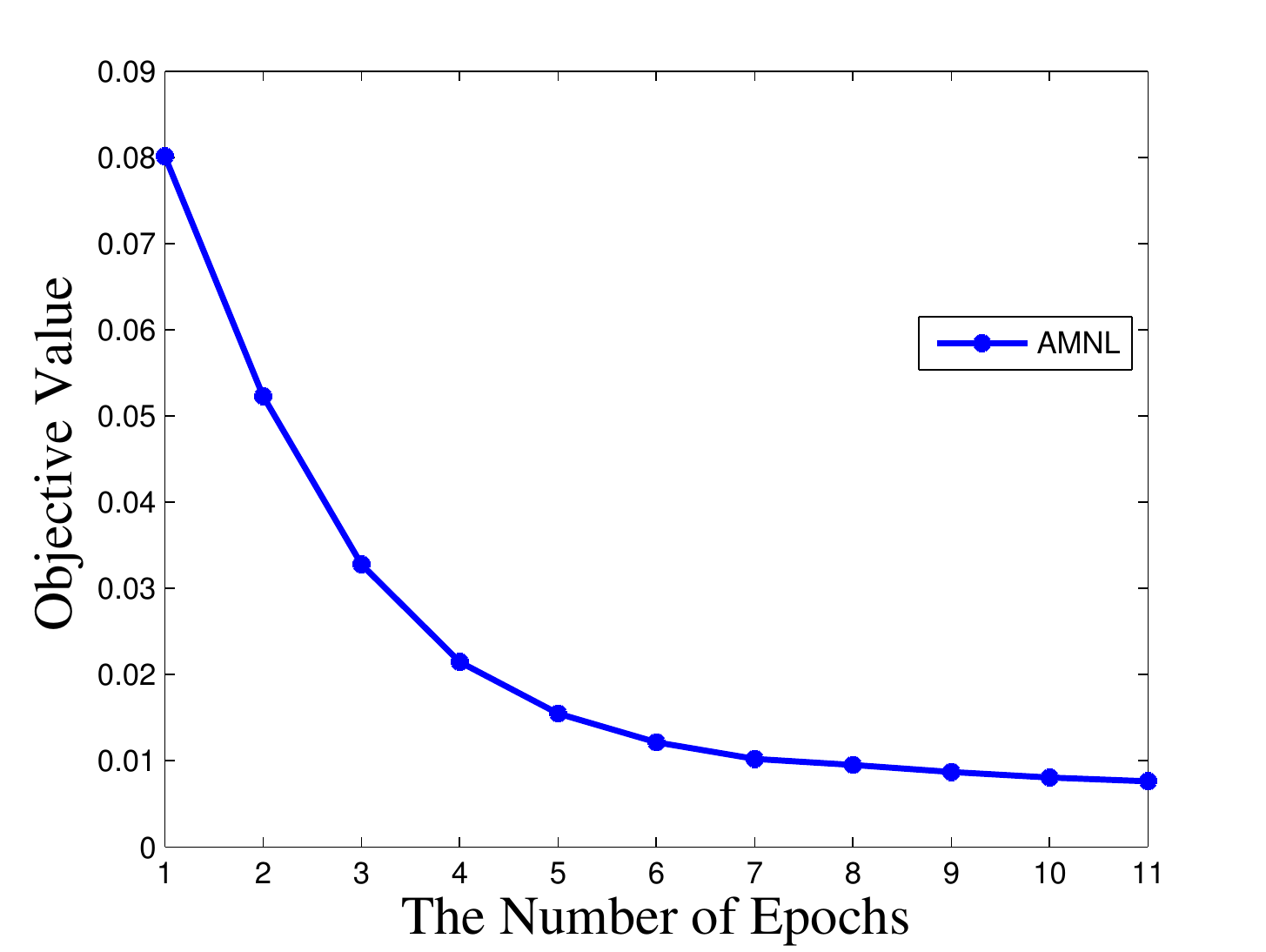}}
	\hspace{.3in}
	\subfloat[Running time]{\includegraphics[width=0.8\columnwidth]{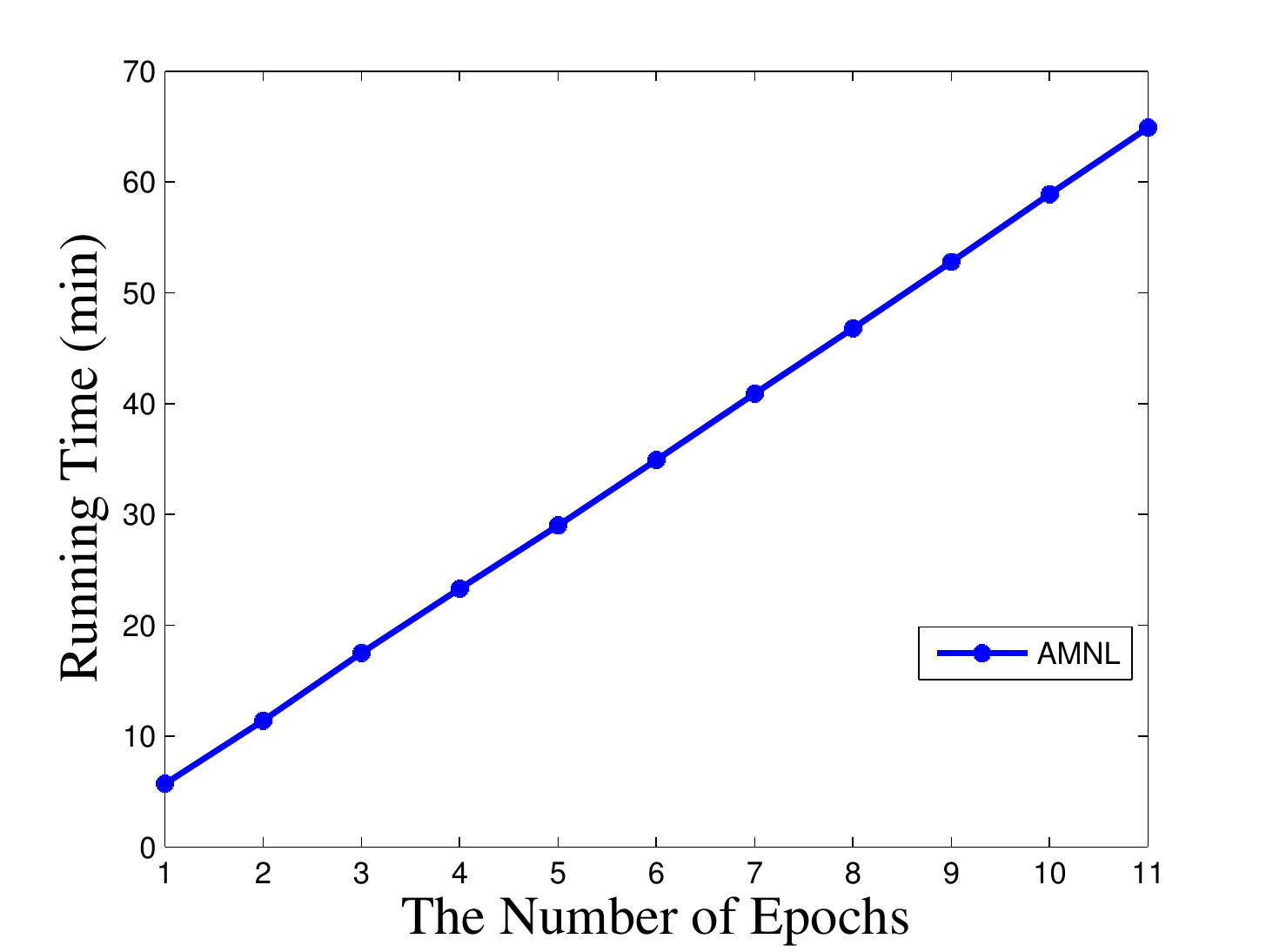}}
	\caption{Objective value and running time versus the number of epochs.}\label{fig:convergence}
\end{figure*}

\begin{figure*}[t]
	\setlength{\belowcaptionskip}{-0.0cm}
	\centering
	\includegraphics[scale=0.8]{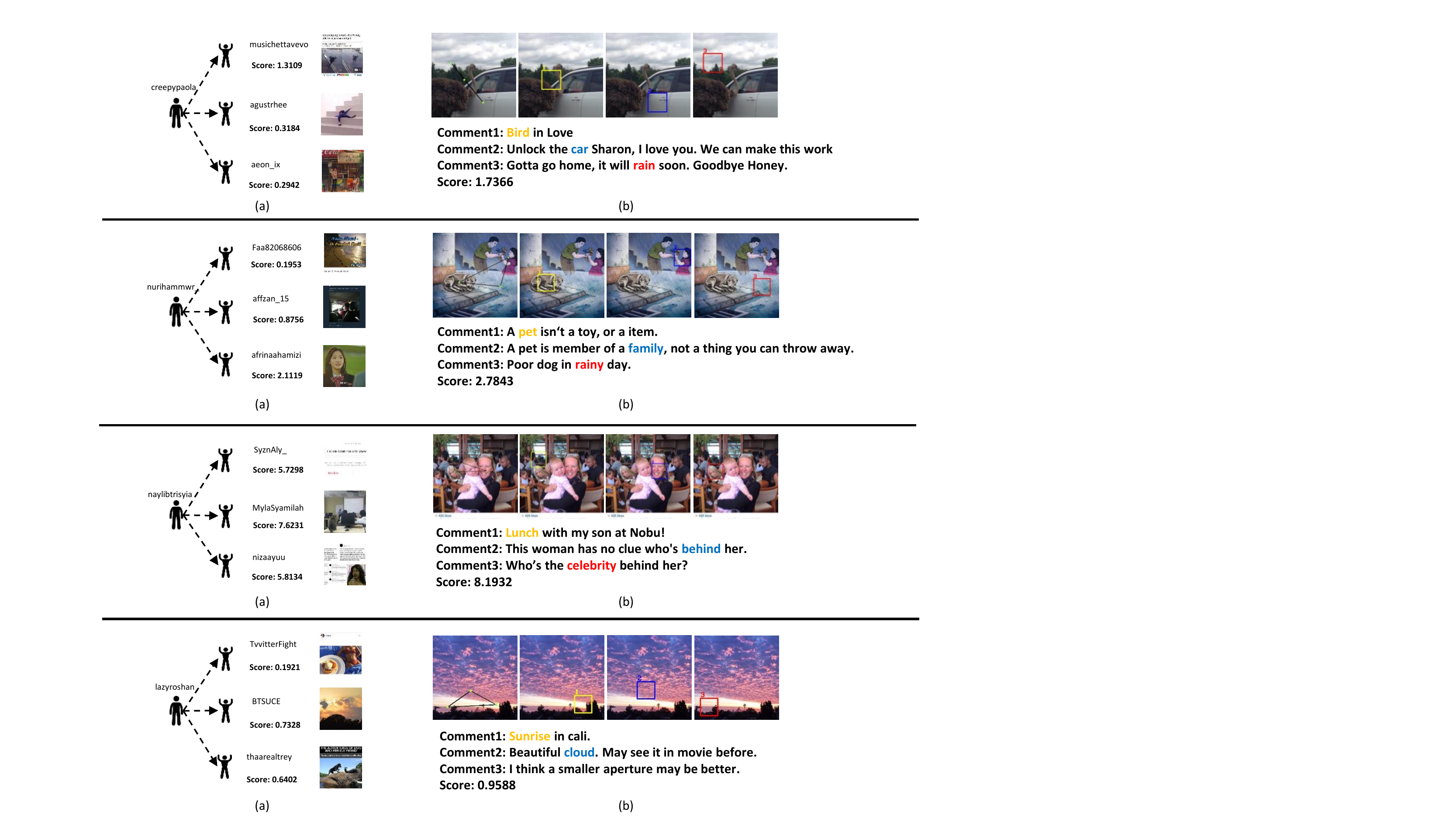}
	\caption{Experimental result of AMNL+ on the image retweet prediction task. (a) The user and unretweeted images published by his/her followees, (b) The preferable retweeted image and its captions or comments predicted by our method}\label{fig:case}
\end{figure*}

Existing retweet prediction methods are mainly based on low-rank factorized ranking model. Methods FAMF, ADABPR and RRFM learn factorized ranking metric based on pairwise preference constraints. Methods CITING and VBPR are feature-aware factorized ranking algorithms based on pairwise preference constraints and feature of item contents.

We extract feature of item contents as follows. Input words of all textual information are initialized by pre-calculated word embeddings and input visual representation of image tweets are initialized by Inception-Net. Parameters of the neural networks used to get the representations of visual content and textual context are updated during training process. The weights of deep neural networks are randomly initialized by a Gaussian distribution with zero mean in our experiments. Following experimental setting in~\cite{chen2016context,he2015vbpr}, we consider the associated textual contexts as the side information of the method CITING and the visual representation of image tweets as the side information of the method VBPR. The hyper-parameters and parameters which achieve the best performance on the validation set are chosen to conduct the testing evaluation. We set the learning rate to 0.01 for the gradient method. We think the top 3 tweets that users want to retweet can reveal the discriminative characteristics of the tweets that users want to retweet. So we evaluate the ranking performance of all methods on the quality of the top 3 ranked image tweets. In order to show the effectiveness of our textually guided multi-modal fusion, we also evaluate the ranking performance of {\bf AMNL} with the simple fusion method we described above. To exploit the effect of the visual representation of image tweets and the semantic representation of the associated contexts to the performance of our method, we denote our {\bf AMNL} method with visual representation of image tweets only by {\bf AMNL}$_{i}$, our {\bf AMNL} method with semantic representation of the associated contexts only by {\bf AMNL}$_{d}$, and our {\bf AMNL+} method with visual representation of image tweets only by {\bf AMNL+}$_{i}$

Tables~\ref{table:precision1},~\ref{table:precision3} and~\ref{table:auc} show evaluation results of all methods on ranking criteria Precision@1, Precision@3 and AUC, respectively. Evaluation were conducted with different ratio of data as training set from 60\%, 70\% to 80\%. We report result value of all methods using three ranking evaluation criteria. We then report performance of our model with different modalities, where dimension of user preference representation is set to 400, and 80\% of data is used for training. All other parameters and hyperparameters are also chosen to guarantee the best performance on the validation set. We evaluate the average value of all three criteria on six methods. These experimental results reveal a number of interesting points:

\begin{itemize}
	\item The methods with content feature as the side information for learning the ranking metric, CITING and VBPR, outperform the low-rank factorized ranking metric methods FAMF, ADABPR and RRFM, which suggests that the deep neural networks with both image tweets and the associated context information is critical for the problem of image retweet prediction.
	\item Compared with other ranking methods with the side information, our method AMNL$_{i}$ achieves better performance than the method VBPR, and our method AMNL$_{d}$ achieves better performance than the method CITING, respectively. This suggests that the multi-faceted ranking metric is important for the problem.
	\item Compared with our methods AMNL, our method AMNL+ achieves better performance. This suggests that through the textually guided multi-modal fusion method, image tweets can be better jointly represented with different captions or comments  which contain the associated semantic information, thus obtaining better performance in the image retweet prediction.
	\item  In all cases, our AMNL+ method achieves the best performance. This shows that the attentional multi-faceted ranking network learning framework that exploits both the joint image tweet representation of multi-modal image tweets and their associated contexts, and multi-faceted ranking metric can further improve the performance of image retweet prediction.
\end{itemize}

We also illustrate the experiment results of our AMNL+ on some users' image retweet prediction in Figure~\ref{fig:case}(a) and (b). The Figure~\ref{fig:case}(a) shows the user and the images published by the user's followees. Their low ranking scores indicate that the nonretweeted image tweets published by followees are more likely to be seen but disliked by the user. The Figure~\ref{fig:case}(b) shows the predicted image and its comments which has a high score. This suggests that the image predicted by our method is more preferable for the user in Figure~\ref{fig:case}(a). It's also worth mentioning that some specific words are matched with objects  marked by the same color in the image, which shows a great effectiveness of the guidance of comments and captions.

\subsection{Hyper-Parameter Analysis}

In our approach, there are three essential parameters, which are the dimension of user preference representation, the dimension of recurrent neural network units and the margin $c$ in the loss function. In order to study the effect of such hyper-parameters, we vary the dimension of user preference representation from 100 to 500, the dimension of recurrent neural network units from 200 to 1200 and the margin value $c$ in the loss function from 0.1 to 0.9. We show the effect of these hyper-parameters using 60\% of the data for training on Precision@1, Precision@3 and AUC in Figures~\ref{fig:dimension06}(a),~\ref{fig:dimension06}(b) and~\ref{fig:dimension06}(c). As is shown in the figures, the change of parameters has a relatively stable effect on the performance of model and the variation tendency is the same. We also find out that with the change of the dimension of user preference representation, all three criteria changes in a larger range than the other two hyper-parameters, which indicates that the dimension of user preference representation is essential for users' image retweet prediction. Our method achieves best performance  when the dimension of user preference representation is set to 400, the dimension of recurrent neural network units is set to 1000 and the margin $c$ in the loss function is set to 0.6 with different proportions of data for training.

The updating rule for training our proposed attentional multi-faceted ranking network learning method is essentially iterative. Here we investigate how our AMNL method converges. Figures~\ref{fig:convergence}(a) and~\ref{fig:convergence}(b) show the convergence and running time curves of AMNL method, respectively. The $x$-axis denotes the iteration number in both figures. The $y$-axis in Figure~\ref{fig:convergence}(a) denotes the objective value and the $y$-axis in Figure~\ref{fig:convergence}(b) shows the running time of our proposed method. Each epoch contains 22,881 iterative updates. We set the dimension of user preference representation to 400, and use 80\% of the data for training. We show that our method converges after 9-th epoch and the computation cost is less than 50 minutes.  This study validates the efficiency of our method.

\subsection{Ablation Study}

In this part, we evaluate the contribution of our technical components: the textually guided multi-modal fusion network and the social impact function. We also evaluate the effect of visual representation of image tweets, semantic representation of the associated contexts and the joint image tweet representation to our model.

To understand the contribution of components and the effect of different media for our model, we propose the ablation study and illustrate the results in Table~\ref{table:modalities}. We explore our model in these ways: our {\bf AMNL}$_{i}$ method means that we use the visual representation of image tweets only. Our {\bf AMNL}$_{d}$ method means that we  only semantic representation of the associated contexts. Our {\bf AMNL+}$_{i}$ model means that we input the average pooling of convolutional feature of image tweets directly into recurrent neural networks in the textually guided multi-modal fusion network, instead of using attention mechanism with the textual representation. Our {\bf AMNL}$_{hfunc}$ and {\bf AMNL+}$_{hfunc}$ model means that we calulate the ranking function directly for two models without using the social impact function. As is shown in Table~\ref{table:modalities}, we also find some interesting results.

\begin{itemize}
	\item Compared with our methods AMNL$_{i}$ and AMNL$_{d}$, our method AMNL achieves better performance. This suggests that the attentional multi-faceted ranking network learning framework which exploits the joint image tweet representation of multi-modal image tweets and their associated context can get better performance than the attentional multi-faceted ranking network learning framework which only exploits the representation of tweets' images or the representation of tweets' contexts.
	\item Compared with result of {\bf AMNL+}$_{hfunc}$, {\bf AMNL+} gets better score among all three criteria. This suggests that the social impact function can help improve the performance of our method. The experiment results of {\bf AMNL}$_{hfunc}$ and {\bf AMNL} further proves that our above result is consistent among different components.
\end{itemize}

\section{Related Work}\label{sec:related work}

Retweet prediction has been studied deeply and extensively in recent years. It's a method to perform information dissemination for today's social media. In order to model user's retweet behavior accurately, we divide the current research work into three aspects: feature selection for user retweet behavior, representation for retweet modeling and user retweet ranking. In this section, we briefly review some related work in all three aspects.

\subsection{Feature Selection of User Retweet Behavior}

How to choose the relevant factor that affect user's retweet behavior has been well studied. ~\cite{Xu2012AnalyzingUR} examines four types of features which are related to the retweetability of each tweet by training a prediction model. ~\cite{Suh2010WantTB} collects both content and contextual features from Twitter dataset and evaluates their affect for retweet behavior. The experiment indicates the great contribution of contextual features to the retweet rate, while the distribution of past tweets does not influence the user's retweetability. ~\cite{Yang2015RAINSR} integrates the social role recognition and information diffusion into a whole framework, modeling the interplay of user's social roles. ~\cite{Kouloumpis2011TwitterSA} examines a number of semantic features to learn the tweets's sentiment representation. ~\cite{Macskassy2011WhyDP} explains that user retweet behavior can be better understood in the unfamiliar area by assessing different predictive models and features. ~\cite{Xu2012ModelingUP} studies the factor of user posting behavior, which consists of breaking news, posts from user's social friends and user's intrinsic interest. The authors also present a latent model to further prove the effectiveness of these factors. ~\cite{Hu2012LearningTS} models both the user's social relation and other factors to perform the retweet prediction. In addition to that, the authors also take the extent difference of social correlation into consideration by dividing them into different categories, such as friends or co-workers. Different from existing methods, our method gathers image tweets and their captions or comments. We suppose that different captions or comments not only represent extensive semantic information for the image, but also have correlation with each other because of the user's socical interaction.

\subsection{Representation for Retweet Modeling}

There has been a number of studies aiming at modeling user's retweet representation. ~\cite{Petrovic2011RTTW} predicts the human retweet behavior by a machine learning approach based on the  passive-aggressive algorithm. ~\cite{Luo2013WhoWR} develops a learning to-rank framework to explore various retweet features. ~\cite{Bourigault2014LearningSN} considers about the task from the perspective of temporal information diffusion. The model learns a diffusion kernel in which the infection time in cascades is represented by the distance of nodes in the projection space. ~\cite{hong2013co} proposes a factorization machine with a ranking-based function, which is extended from a recommendation model, to integrate various aspects in Twitter dataset. ~\cite{Boyd2010TweetTR} converts the task of retweet modeling into the  conversational practice, in which the authorship and communicative fidelity are negotiated. ~\cite{Lim2013RetweetingAA} treats the retweet behavior as a three-dimensional tensor of tweets, tweet authors and their followers and represents them simultaneously by tensor factorization. ~\cite{Matsubara2012FastMA} collects the interplay of users and contextual information, using a support vector data description to predict the future interplay.  ~\cite{Jiang:2015:MCB:2806416.2806650} deploys the matrix completion approach to optimize the factorization of user's retweet representation. Despite that previous studies have explored a wide range of representation learning for the user's retweet modeling, most of them do not specifically take account of the jointly representation of image retweets and their captions or comments, for which we propose the textually guided multi-modal network and evaluate its effectiveness using Twitter dataset.

\subsection{User Retweet Ranking}

Central problem of retweet prediction is to model tweet sharing behavior that users repost tweets along followee-follower links and rank all tweets emerged in social media so that more users are informed in SMS, which has attracted considerable attention recently in~\cite{chen2016context,firdaus2016retweet,zhang2015retweet,zhang2016retweet,wang2013whom,feng2013retweet}. Chen et. al.~\cite{chen2016context} exploit various contexts for image understanding and retweet prediction. Firdaus et. al.~\cite{firdaus2016retweet} propose a retweet prediction model by considering user's author and retweet behaviors. Zhang et. al.~\cite{zhang2015retweet} propose non-parametric models to combine structural, textual, and temporal information together to predict retweet behavior. Zhang et. al.~\cite{zhang2016retweet} propose deep neural networks to incorporate contextual and social information. Wang et. al.~\cite{wang2013whom} present a recommendation model to solve the problem of whom to mention in a tweet. Feng et. al.~\cite{feng2013retweet} propose the feature-aware factorization model to re-rank the tweets, which unifies the linear discriminative model and the low-rank factorization model. Peng et. al.~\cite{peng2011retweet} model the retweet behavior and rank the tweets using conditional random fields. Zhang et. al.~\cite{zhang2015influenced} employ the social influence locality for ranking the user's retweets rate. Unlike previous studies, we formulate the problem of image retweet prediction from the viewpoint of attentional multi-faceted ranking network learning, which can be solved by the negative sample based ranking metric learning with multi-modal neural networks.

\section{Conclusion}\label{sec:conclusion}

In this paper, we introduced problem of image retweet prediction from viewpoint of attentional multi-faceted ranking network learning. We propose heterogeneous IRM network that exploits both users' past retweeted image tweets, associated textual context and users' following relations. We present a novel attentional multi-faceted ranking network learning method with the textually guided multi-modal neural networks to learn joint image tweet representations and user preference representations, such that multi-faceted ranking metric is embedded in representations for prediction. We evaluate performance of our method using dataset from Twitter. Extensive experiments demonstrate that our method can achieve better performance than several state-of-the-art solutions.

% or


%\appendices

%\section{}

%\section*{Acknowledgment}



%\ifCLASSOPTIONcaptionsoff
%  \newpage
%\fi



\bibliographystyle{IEEEtran}
\bibliography{AMNLijcai18}

%\begin{thebibliography}{1}


%\end{thebibliography}


\begin{IEEEbiography}[{\includegraphics[width=1in,height=1.25in,clip,keepaspectratio]{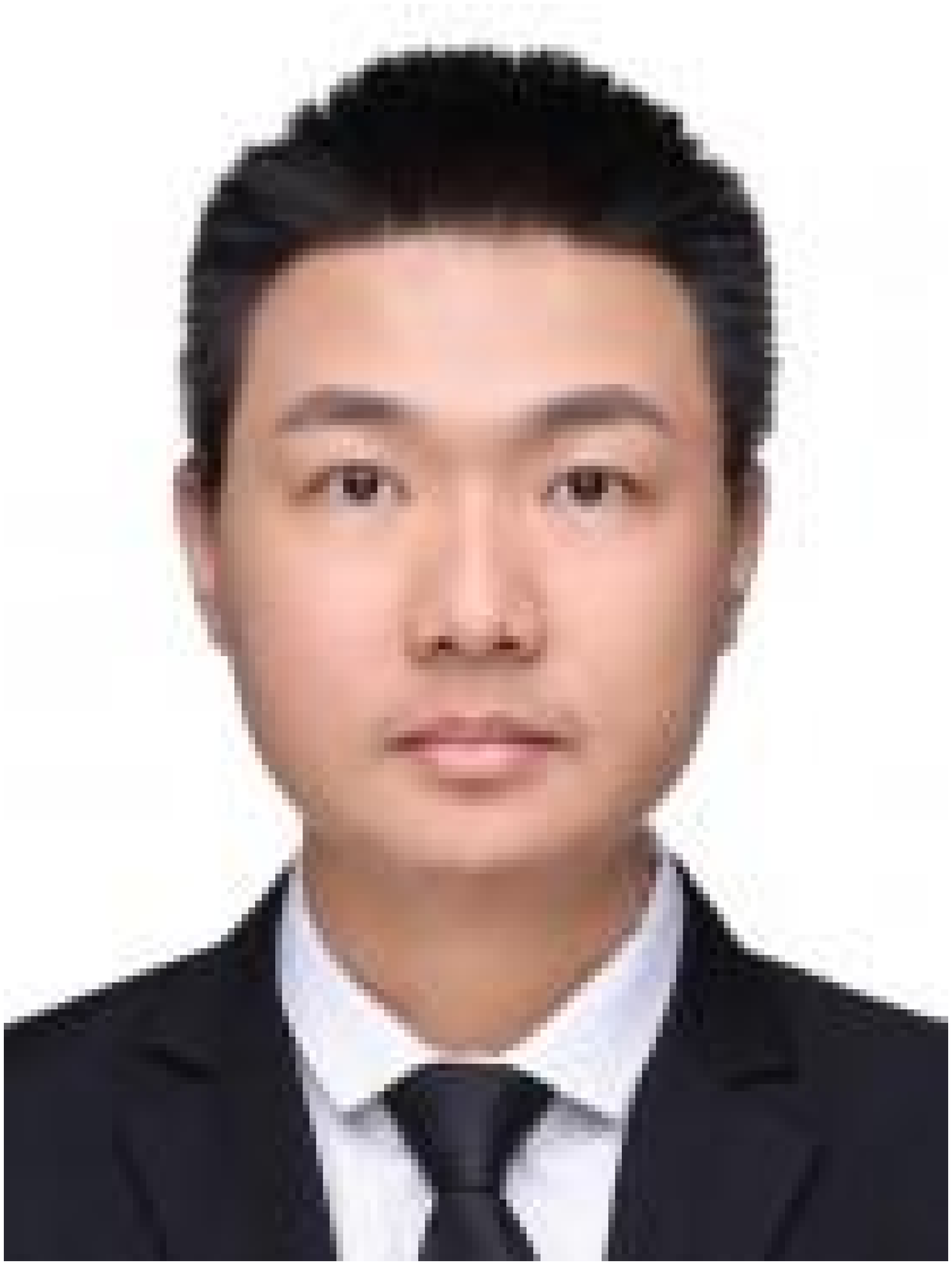}}]{Zhou Zhao}
	received the B.S and Ph.D. degrees in computer science from The Hong Kong University of Science and Technology, in 2010 and 2015, respectively. He is currently an Associate Professor with the College of Computer Science, Zhejiang University. His research interests include machine learning and data mining
\end{IEEEbiography}

\begin{IEEEbiography}[{\includegraphics[width=1in,height=1.25in,clip,keepaspectratio]{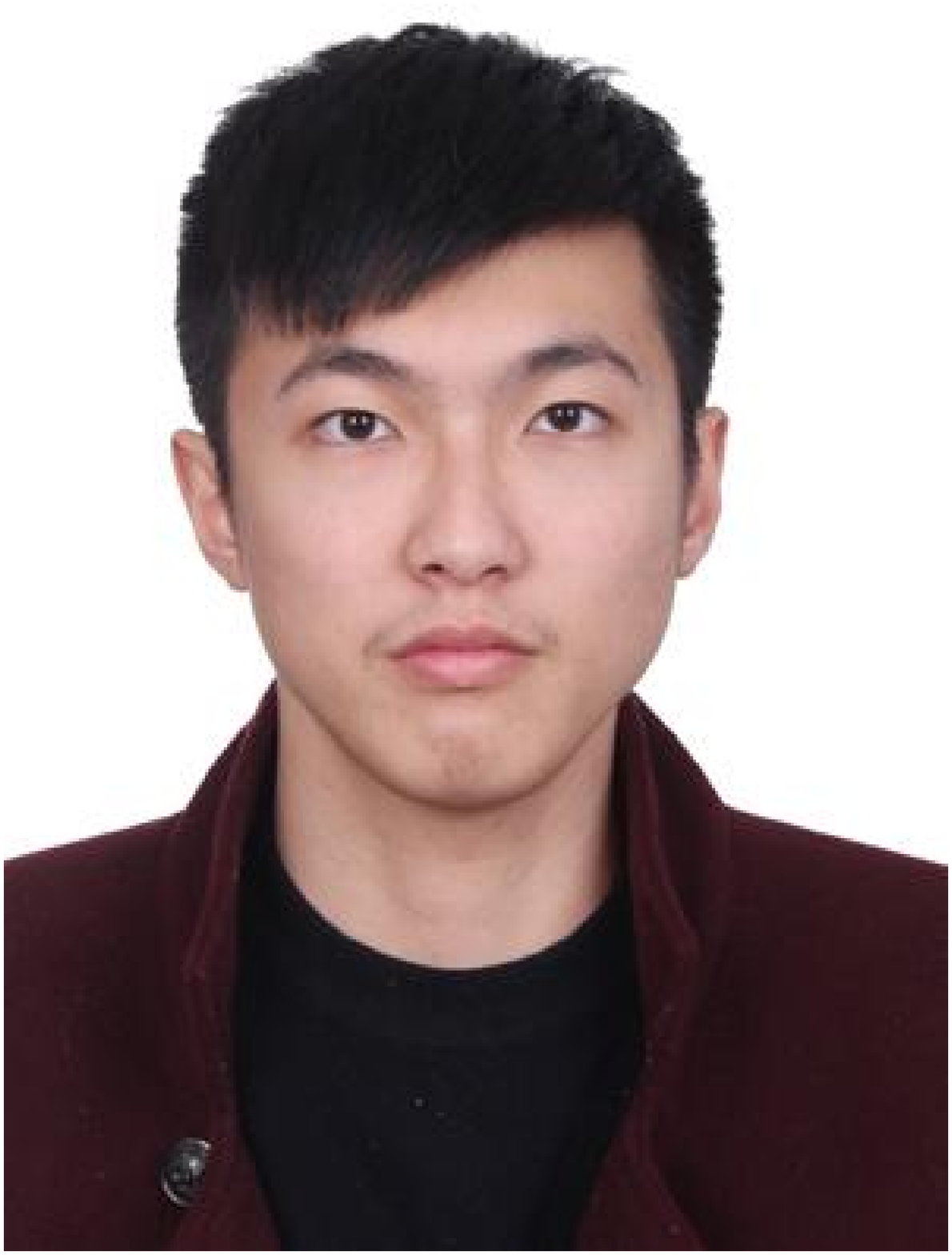}}]{Hanbing Zhan}
	is currently pursuing the B.E. degree in computer science and technology from Zhejiang University. His research interests include multimedia analysis, computer vision and natural language processing.
	
\end{IEEEbiography}

\begin{IEEEbiography}[{\includegraphics[width=1in,height=1.25in,clip,keepaspectratio]{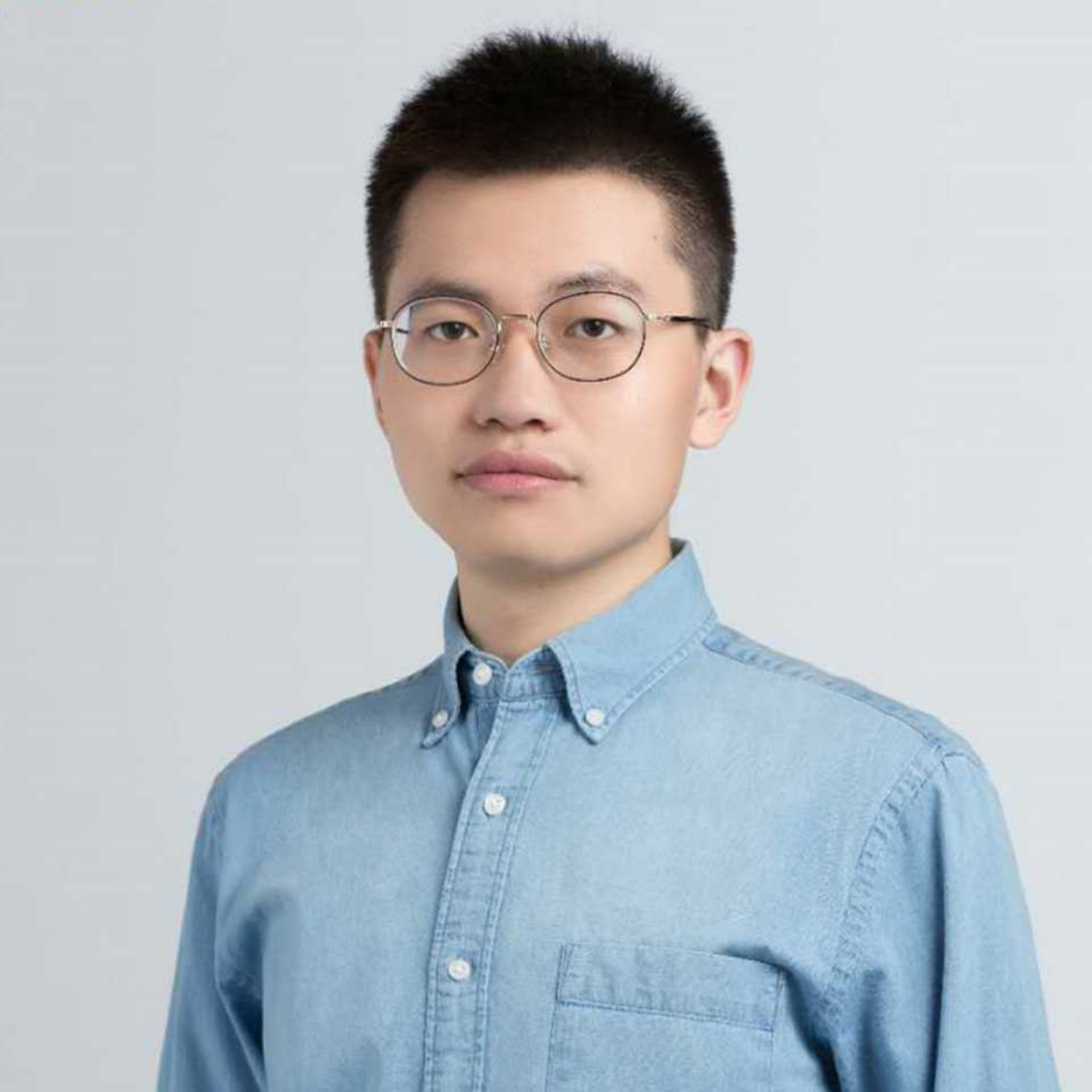}}]{Lingtao Meng}
	received the B.E. degree in software engineering from Shandong University,
	China, in 2016. He is currently pursuing the M.S. degree in computer science from Zhejiang University. His research interests include machine learning, computer vision and natural language processing.
\end{IEEEbiography}

\begin{IEEEbiography}[{\includegraphics[width=1in,height=1.25in,clip,keepaspectratio]{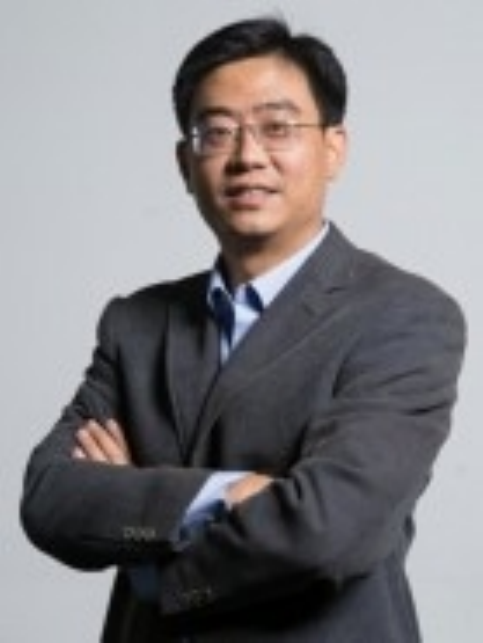}}]{Jun Xiao}
	is currently a Full Professor with the College of Computer Science, Zhejiang University. He received the B.E. and Ph.D. degree from the College of Computer Science, Zhejiang University, Hangzhou, China, in 2002 and 2007, respectively. His research interests include cross-media analysis, computer vision, and machine learning.
\end{IEEEbiography}

\begin{IEEEbiography}[{\includegraphics[width=1in,height=1.25in,clip,keepaspectratio]{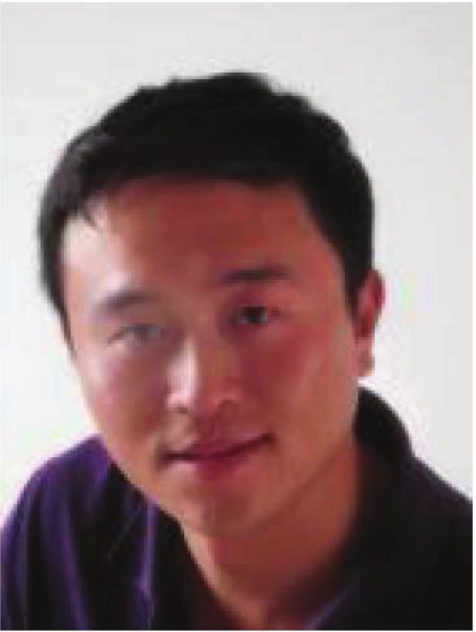}}]{Jun Yu}
	received the B.Eng. and Ph.D. degrees from Zhejiang University, Zhejiang, China. He is currently a Professor with the School of Computer Science and Technology, Hangzhou Dianzi University, Hangzhou, China. He was an Associate Professor with the School of Information Science and Technology, Xiamen University, Xiamen, China. From 2009 to 2011, he was with Nanyang Technological University, Singapore. From 2012 to 2013, he was a Visiting Researcher at Microsoft Research Asia (MSRA). Over the past years, his research interests have included multimedia analysis, machine learning, and image processing. He has authored or coauthored more than 80 scientific articles. In 2017 Prof. Yu received the IEEE SPS Best Paper Award.
	Prof. Yu has (co-)chaired several special sessions, invited sessions, and workshops. He served as a program committee member or reviewer of top conferences and prestigious journals. He is a Professional Member of the Association for Computing Machinery (ACM) and the China Computer Federation (CCF).

\end{IEEEbiography}
\newpage
\begin{IEEEbiography}[{\includegraphics[width=1in,height=1.25in,clip,keepaspectratio]{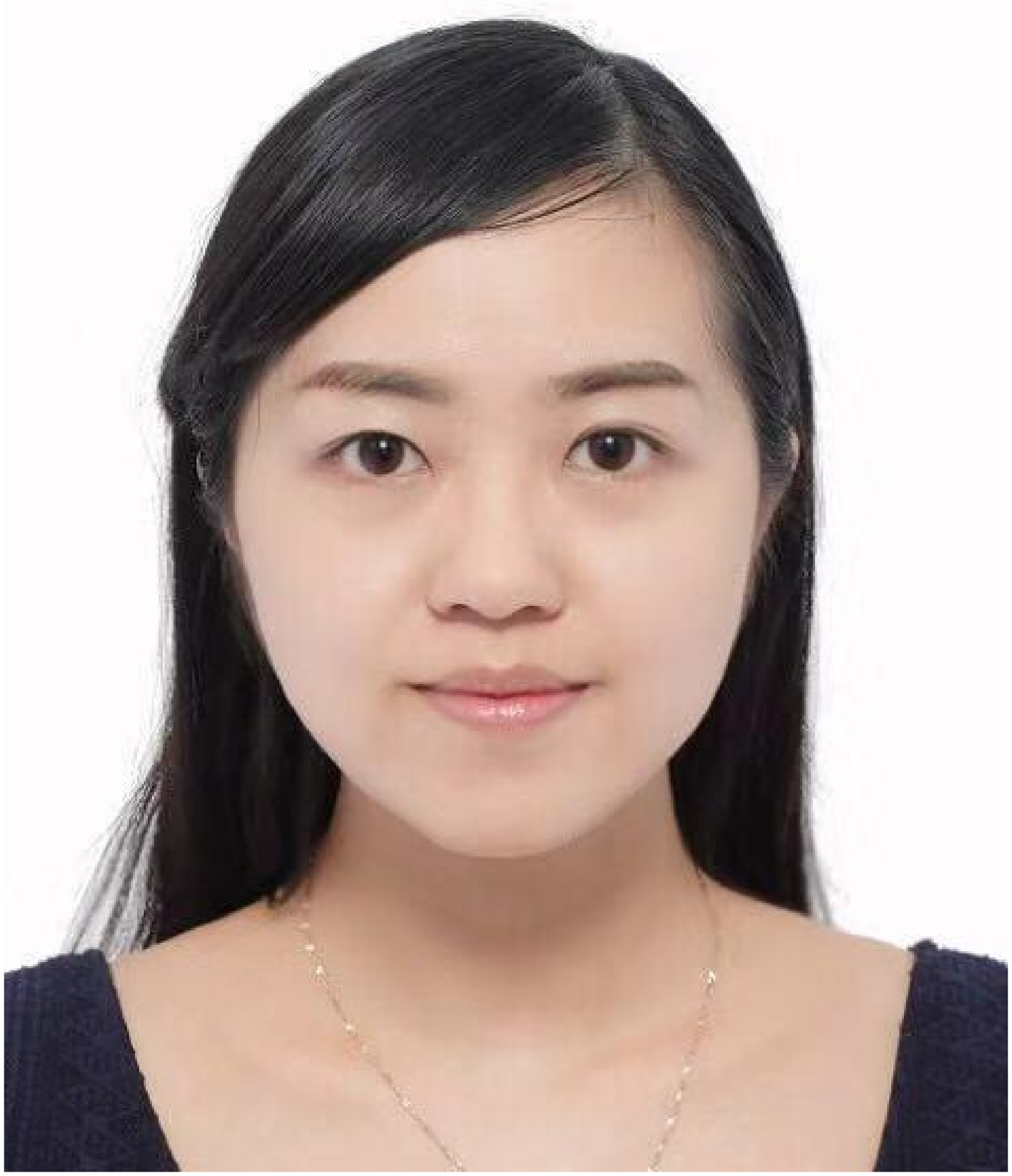}}]{Min Yang}
	is now an Assistant Professor in Shenzhen Institutes of Advanced Technology (SIAT), Chinese Academy of Sciences, leading the Artificial Intelligence and Natural Language Processing team in Global Center for Big Mobile Intelligence. Previously, she was a senior AI researcher at Tencent, working on NLP related tasks. She received her Ph.D. degree in Department of Computer Science, The University of Hong Kong, advised by Dr. K.P. Chow. Before that, she got her Bachelor degree from Sichuan University, 2012. She was a visiting student with Rensselaer Polychnic Institute from March 2016 to June 2016. Her research interests include a variety of topics, including Natural Language Processing (e.g., sentiment analysis, topic modeling, event extraction, question answering, knowledge base, named entity recognition, user profiling, etc.), Multimodal Learning (e.g., image captioning, video summarization), Deep Learning, Recommendation Systems, Machine Learning in Finance.
	
\end{IEEEbiography}

\begin{IEEEbiography}[{\includegraphics[width=1in,height=1.25in,clip,keepaspectratio]{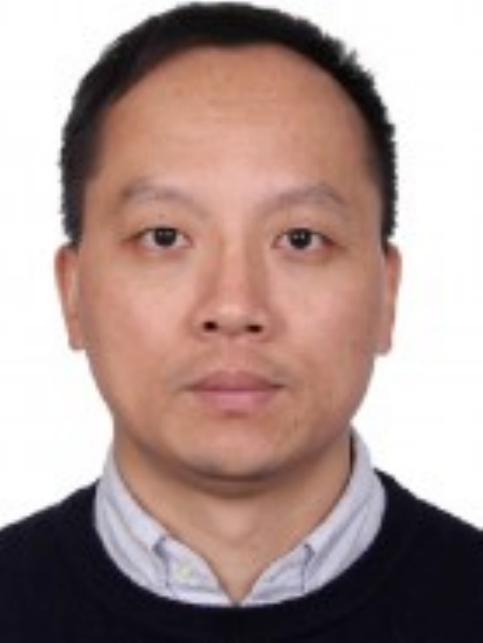}}]{Fei Wu}
	received his B.Sc., M.Sc. and Ph.D. degrees in computer science from Lanzhou University, University of Macau and Zhejiang University in 1996, 1999 and 2002 respectively. From October, 2009 to August 2010, Fei Wu was a visiting scholar at Prof. Bin Yu's group, University of California, Berkeley. Currently, He is a Qiushi distinguished professor of Zhejiang University at the college of computer science. He is the vice-dean of college of computer science, and the director of Institute of Artificial Intelligence of Zhejiang University. He is currently the Associate Editor of Multimedia System, the editorial members of Frontiers of Information Technology \& Electronic Engineering. He has won various honors such as the Award of National Science Fund for Distinguished Young Scholars of China (2016). His research interests mainly include Artificial Intelligence, Multimedia Analysis and Retrieval and Machine Learning.
\end{IEEEbiography}

\begin{IEEEbiography}[{\includegraphics[width=1in,height=1.25in,clip,keepaspectratio]{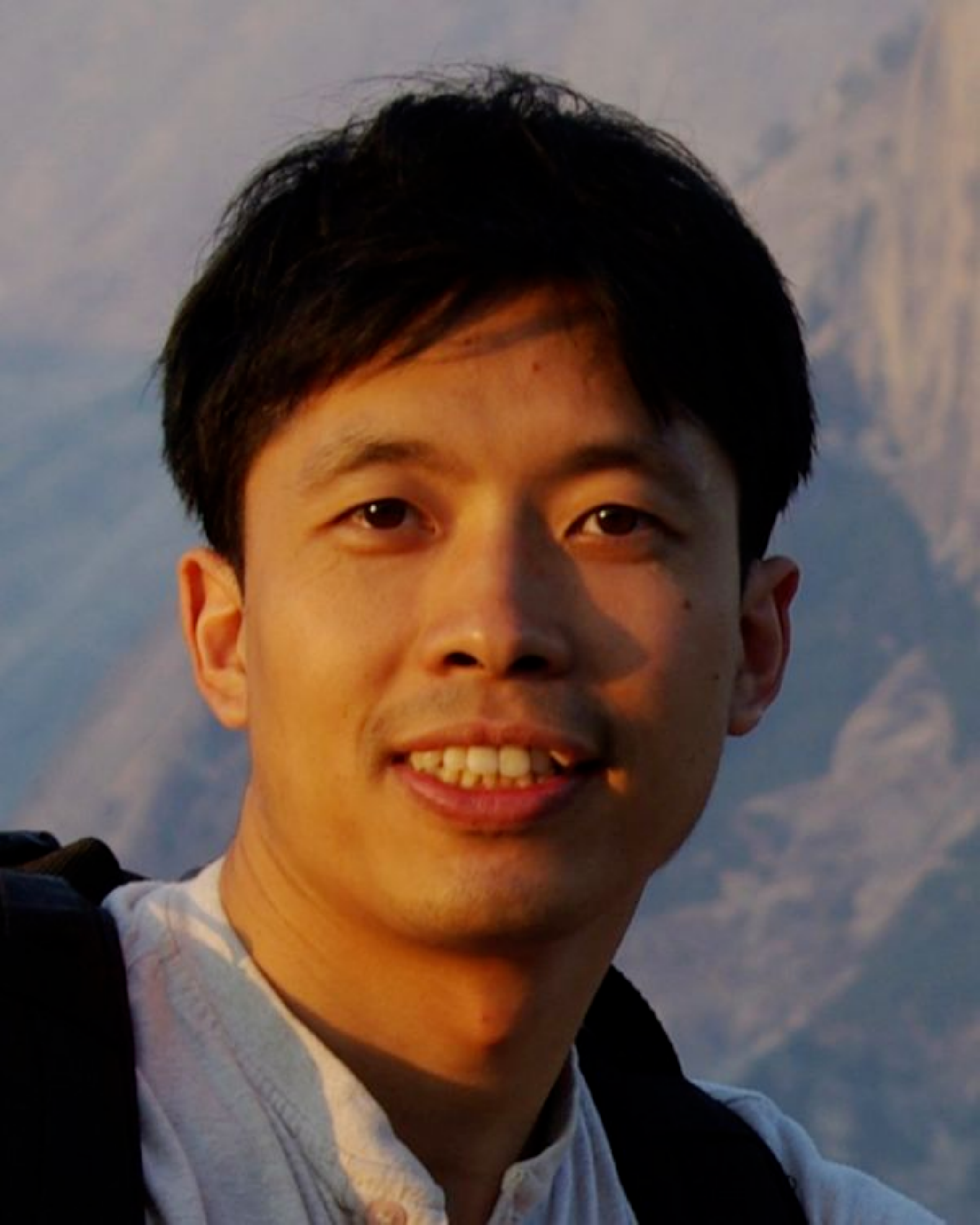}}]{Deng Cai}
	is currently a full professor in the College of Computer Science at Zhejiang University, China. He received the PhD degree from University of Illinois at Urbana Champaign. His research interests include machine learning, computer vision, data mining and information retrieval.
\end{IEEEbiography}


%\newpage



%\vfill


\end{document}